\documentclass[nofootinbib,notitlepage,onecolumn]{revtex4-1}
\setcitestyle{super}

\pdfpagewidth=\paperwidth
\pdfpageheight=\paperheight

\usepackage{amsmath,graphicx}
\usepackage{verbatim}
\usepackage{color}
\usepackage{amsfonts,amsbsy,amssymb}
\usepackage{jabbrv}
\usepackage{natmove}
\usepackage{natbib}
\makeatletter
\def\tagform@#1{\maketag@@@{[\ignorespaces#1\unskip\@@italiccorr]}}
\makeatother

\usepackage{titlecaps}
\Addlcwords{the of into and in a with from on its ab initio to for}

\usepackage{cancel}
\newcommand{\changeAD}[1]{\textcolor{black}{#1}}
\newcommand{\changepp}[1]{\textcolor{black}{#1}}

\newcounter{paulfoot}

\makeatletter
\def\blfootnote{\stepcounter{paulfoot} (see footnote \arabic{paulfoot})\gdef\@thefnmark{\arabic{paulfoot}}\@footnotetext}
\makeatother

\newcommand{\ocite}[1]{[\hspace*{-1mm}\citenum{#1}]}

\def \pdf {\mathcal P}

\def \bq {\boldsymbol {\rm q}}
\def \bk {\boldsymbol {\rm k}}
\def \by {\boldsymbol {\rm Y}}
\def \bp {\boldsymbol {\rm p}}
\def \bx {\boldsymbol {\rm X}}
\def \bF {\boldsymbol {\rm F}}

\def \bHam {\boldsymbol {\rm H}}
\def \bz {z}

\def \massM {\boldsymbol{ \mathcal M}}
\def \Dh {{\mathcal D}_h}
\def \Fh {{\mathcal F}_h}

\def \amBar {\bar{a}_m}

\def \bbq {\boldsymbol {\rm Q}}
\def \bbp {\boldsymbol {\rm P}}
\def \b1 {\boldsymbol {\rm 1}}
\def \bzero {\boldsymbol {\rm 0}}
\def \bj {\boldsymbol {\rm J}}

\def \d  {{\rm d}}
\def \dt {{\rm d} t}
\def \dbx {{\rm d} \bx}

\def \gradH  {\nabla H}
\def \gradHP {\nabla_{\bbp} H}
\def \gradHQ {\nabla_{\bbq} H}
\def \dotQP {\begin{pmatrix} \dot \bbq \\ \dot \bbp \end{pmatrix}}

\def \bM {\boldsymbol {\rm M}}
\def \bMRK {\bM_{\rm RK4}}
\def \bMVV {\bM_{\rm VV}}

\newcommand{\matr}[4]{\begin{pmatrix} #1 & #2 \\ #3 & #4 \end{pmatrix}}

\def\e{\rm e}
\def\tg{\ensuremath{T_{\rm g}}}
\def\Tg{\ensuremath{T_{\rm g}}~}

\def\TgBar{\ensuremath{\bar T_{\rm g}}}
\def\TgBari{\ensuremath{\hat T_{{\rm g},i}}}
\def\tgi{\ensuremath{T_{{\rm g},i}}}
\def\E{\boldsymbol {\rm E}}

\def\T{\boldsymbol {\rm T}}
\def\d{~\rm d}

\def\rhoH{\rho_\H}
\def\H{\mathcal H}
\def\brho{\boldsymbol {\rho}}

\def\Var{{\rm Var}}

\def\TTg{{\rm \bf \hat T_{\rm g}}}

\def\brhowithin{\boldsymbol {\omega}_i}

\def\tgnoise{\tau}
\def\tgwithin{\tgnoise_{\omega,i}}
\def\tgbetween{\tgnoise_{\beta,i}}
\DeclareMathOperator{\Cov}{Cov}
\DeclareMathOperator{\argmin}{argmin}

\newcounter{theorem}
\newtheorem{thm}[theorem]{Theorem}

\begin{document}

\author{Paul N. Patrone}
\email{paul.patrone@nist.gov}
\author{Andrew Dienstfrey}
\email{andrew.dienstfrey@nist.gov}
\affiliation{National Institute of Standards and Technology\footnote{This work is a contribution of the National Institute of Standards and Technology and is not subject to copyright in the United States.}}

\date{\today}
\title{Uncertainty Quantification for Molecular Dynamics}

\maketitle

\section*{Introduction}

Since the 1960's, scientists have increasingly turned to molecular dynamics (MD) as a tool of choice for studying complex and realistic condensed-matter systems.  The reason behind this trend is clear: computers can solve high-dimensional mathematical problems that are otherwise intractable \cite{Adler59}.  With the growing availability of high-performance computers, MD simulations have therefore  allowed scientists to model systems of increasing size and complexity,  opening up the possibility of understanding physical phenomena across multiple scales  \cite{Perilla15,Shibuta14}.  In recent years, some studies have even had success bridging all the way from atomic to thermodynamic scales, modeling the properties of macroscopic materials in terms of their microscopic components \cite{Strachanreview,stevepat1,stevepat2}.

While this scientific rationale has driven many past efforts, more pragmatic motivations are beginning to take hold in the community and shape the way simulations are used.  Of note, the advent of readily available and easy-to-use software packages has led to the perception that MD can be treated as a black box \cite{Blackbox1,BB2,BB3,BB4}.  In and of itself, this is not entirely bad; by hiding the details of every subroutine and computation, such tools essentially filter out information that is not easily interpreted by non-specialists.  Moreover, these packages have made MD accessible to a wider audience by decreasing the technical overhead needed to train new modelers \cite{economics}.  The benefits of this accessibility have been so dramatic that computational tools now occupy a variety of positions in industrial ${\rm R\& D}$ settings and have been attributed, for example, with speeding up product development in the aerospace community \cite{ICME1,ICME2,stevepat1,stevepat2,economics}.  

Despite these benefits, the ``black-box'' mentality can also introduce drawbacks.  In particular, the availability of easy-to-use MD packages has sometimes led to the misconception that a result is correct because the simulation did not crash.  Given the complexity of the MD enterprise, however, it can be difficult for even experts to assess the quality of simulated predictions \cite{binder95}.  Moreover, this problem stands to worsen as MD gains traction in scientific communities that are not intimately familiar with its underlying mathematical and physical assumptions.  As a result, computational scientists are beginning to recognize that systematic techniques for assessing the reliability of MD computations are needed for this tool to be leveraged more widely and with greater confidence \cite{ICMENRC,IMAJOM}. We use the term {\it uncertainty quantification} (UQ) to refer to the collection of techniques whose overarching objective is to increase such confidence\blfootnote{We note that in a variety of industrial and policy-making contexts, not only molecular dynamics but scientific computing more generally is increasingly being used to inform costly and consequential decisions.  As a result, a growing number of stakeholders now view uncertainty quantification as a necessary component of decision-making workflows informed by simulation \cite{economics,ICMENRC}.}.

In light of these observations, the goals of this chapter are therefore twofold.  First, we wish to introduce MD and UQ in a common setting in order to demonstrate how the latter can increase confidence in the former.  In some cases, this discussion culminates in our providing practical, mathematical tools that can be used to answer the question, ``is this simulation reliable?'' However, many questions remain unanswered.  Thus, a second goal of this work is to highlight open problems where progress would aid the larger community.  

The UQ problems discussed herein are best understood in the context of MD as a tool for computing the properties of macroscopic equilibrium systems \cite{binder95,Limecooler,allen89}.  In statistical mechanics, such properties are computed as {\it ensemble averages} of the form
\begin{align}
\langle G \rangle = \int {\d} \bx \,\, \pdf(\bx) G(\bx) \label{eq:expectation}
\end{align}
where $\langle G \rangle$ is a {\it macroscopic observable}, $G(\bx)$ is its counterpart in terms of a {\it microstate} $\bx$---i.e., coordinates and momenta of the microscopic particles---and $\pdf(\bx)$ is the probability of the given microstate, which depends on the energy $H(\bx)$ and thermodynamic ensemble. Despite its theoretical simplicity, Eq.~\eqref{eq:expectation} is rarely useful in practice because the complexity of $\pdf(\bx)$ defies analysis.  Molecular dynamics overcomes this issue by replacing the integral with a quadrature estimate of the form
\begin{align}
\langle G \rangle = \int \dbx \,\, \pdf(\bx) G(\bx)\approx \frac{1}{M}\sum_{i=1}^M G(\bx_i) \label{eq:MCapprox}
\end{align}
where the $\bx_i$ are random microstates drawn from the distribution $\pdf(\bx)$ \cite{allen89,Limecooler}.  To compute these $\bx_i$, MD invokes a dynamical algorithm in which the $\bx_i$ are approximations of the trajectory of the molecular system evaluated at the $i$-th timestep according to Newton's equations, i.e.\  $\bx_i\approx\bx(t_i)$.  In this sense, MD can be viewed as a ``deterministic random-number generator,'' an idea that will play a central role in the discussion to come.  

This notion of dynamical sampling brings with it several vexing issues whose resolutions range from subtle to incomplete. For one, while Newtonian mechanics is well-accepted, specification of forces is a critical ingredient in this recipe.  At atomistic scales these are surprisingly complicated  and often involve heuristic approximations that attempt to mimic both classical and quantum effects.  Further issues arise in discrete integration of the Hamiltonian system of ordinary differential equations. In particular, MD systems are chaotic in the sense that two systems initialized arbitrarily closely will quickly diverge from one another in phase space as they evolve forward in time. From the perspective of numerical analysis, this suggests that standard notions of trajectory convergence via step-size refinement are problematic, if not misguided.  Moreover, computational limitations typically restrict the number of particles to a few tens of thousands and integration times to nanoseconds, both of which are far from bulk scales.  In light of this, one may question the degree to which the statistical averaging offered in equation \eqref{eq:MCapprox} represents the desired ensemble average. We will address some of these issues in more detail below.  

In passing, we also mention that the equality of the expectation over $\pdf(\bx)$ with an expectation over time amounts to the statement that the dynamical system is {\it ergodic} \cite{Pathria,landau}.  While it is generally assumed that MD systems have this property, rigorous proofs are hard won \cite{Limecooler}. Keeping with common practice, we assume ergodicity throughout this chapter.  We also note that not all MD studies are designed to extract thermodynamic averages.  One prominent example is the use of MD in protein folding studies, which elicit the kinetic pathways that drive microscopic state changes \cite{folding1,folding2}. Such usages invite their own lines of questioning \cite{Xia:2014,Mor08,Patra203,Cheatham00,Biostuff}, but we do not address them here. 

In light of the issues described above, it is worth emphasizing that our  use of the phrase {\it uncertainty quantification} encompasses a wider range of topics than might appear in more traditional surveys of the subject.  In particular, we adopt the perspective that {\it UQ ultimately provides information for the purposes of making decisions.}  Thus, as the name suggests, we sometimes compute {\it error bars} or {\it confidence intervals} for a simulated prediction. But given that MD is based on well-known mathematical and thermodynamic principles, we also consider consistency checks in order to build confidence that we are modeling the desired physics. In all cases, we offer our UQ approach as a suggested pathway, recognizing that more effective alternatives may emerge in the future.

In our experience, a key barrier to framing UQ of MD is the fact that the relevant topics are dispersed across many, sometimes unconnected fields.  Stove-piping of these disciplines is somewhat inevitable: numerical analysts are primarily concerned with the accuracy of computational algorithms, physicists care about thermodynamic relationships, and statisticians focus on uncertainties associated with data analysis.  Thus, our challenge is to synthesize these ideas in a way that does service to the technical details while also being accessible to computational material scientists, our intended audience.  To this end, we have tried to limit the amount of background material required to understand this chapter.  Most ideas will be introduced as needed, although we assume some familiarity with basic concepts in statistical mechanics (e.g.\ thermodynamic ensembles) and probability theory (e.g.\ random variables, probability densities and distributions).  From a practical standpoint, access to the open-source molecular dynamics package LAMMPS \cite{LAMMPS} is also useful, since the accompanying simulation files are written in that language.  Access to MATLAB \cite{MATLAB} or Octave \cite{octave} is helpful (but not necessary) to run the tutorials, since analysis scripts are provided as down-loadable text files that can be read with one's favorite editor\blfootnote{Certain commercial products are identified in this chapter in order to specify the computational procedure adequately. Such identification is not intended to imply recommendation or endorsement by the National Institute of Standards and Technology, nor is it intended to imply that the materials or equipment identified are necessarily the best available for the purpose.}. Finally, all of the tutorial scripts in this chapter have been uploaded and tested on nanoHUB, a scientific cloud computing service hosted by Purdue \cite{nanohub}. We invite readers to create a nanoHUB account and experiment with scripts there\blfootnote{Scripts are available for download from YYYYYYYY.}.  

The rest of this chapter is laid out as follows.  We begin by covering basic ideas in the theory of molecular dynamics, with the goal of highlighting the main assumptions and underlying issues that point to the need for UQ.  Following this, we provide background on general ideas in uncertainty quantification in an effort to familiarize the reader with concepts that appear both in the tutorials and the broader literature.  Next, we apply these ideas to MD simulations and address the main questions of this chapter by means of simple examples.  We conclude with  final thoughts and open directions.  Throughout, we point the reader to relevant works in the literature that form the basis for this discussion.

\section*{From Dynamical to Random: Overview of MD}
\label{sec:MD}
As discussed in the introduction, molecular dynamics (MD) is founded on the powerful idea that there should be an equivalence between dynamical chaos of many-body systems and the probability distributions predicted by statistical mechanics.  In other words, solving Newton's equations should generate samples $\bx(t)$ for use in Eq.~\eqref{eq:MCapprox}.  However, for many-body systems, it is impossible to solve the corresponding differential equations analytically, and therefore any practical MD algorithm turns to a variety of techniques built upon discrete approximations.  

The purpose of this section is therefore to provide a high-level overview of MD with the goal of highlighting problems that (i) are discussed in the tutorials or (ii) remain open within the field.  As many of these problems amount to confirmation of consistency with general laws from statistical mechanics, we also review relevant aspects of the latter.  For a more complete treatment of MD in the context of statistical mechanics, see Refs.\ \ocite{Limecooler,allen89,binder95}.

The {\it Hamiltonian formulation} of classical mechanics---a succinct and powerful reformulation of Newton's equations---provides the mathematical foundation for MD \cite{Limecooler}. At first glance, it is not obvious in what manner Newton's equations benefit from such a reformulation, especially as the physics is not actually altered.  What we gain, however, is a new perspective that allows for a deeper understanding of their mathematical structure, e.g.\ in terms of invariants. We touch upon these ideas in later sections.  See also Chapters 8-10 of Ref.\ \ocite{goldstein}. A full prescription of a Hamiltonian system requires specification of several items: (i) the system geometry including boundary conditions, constituent particles, and their initial conditions; (ii) interatomic potentials governing the dynamics; (iii) laws of motion and discretization thereof; and (iv) external thermodynamic parameters such as the temperature or pressure.  We describe these elements in more detail below.

\subsection*{System Specification}
\label{subsec:geometry}
\begin{figure}
\includegraphics[width=16cm]{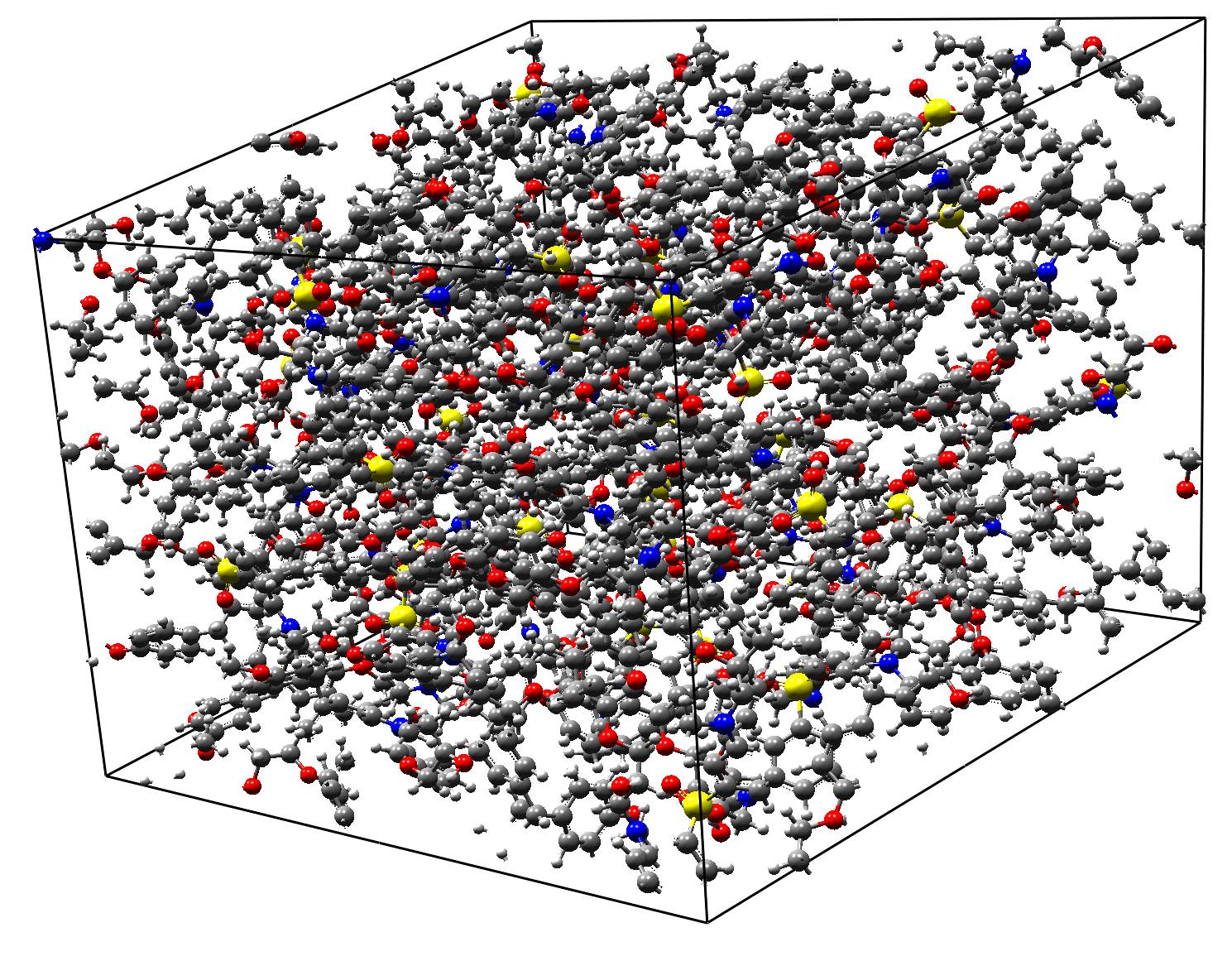}\caption{An example of a periodic unit cell.  Particles that leave the cell through one boundary re-enter from the opposite side.  The number of atoms in the figure (roughly 4000) is representative of a typical system that one might model in a high-throughput industrial setting.}\label{fig:unitcell}
\end{figure}

The system geometry (or domain $D$) is generally an orthorhombic {\it unit cell}; see Fig.~\ref{fig:unitcell}.  Intuitively, this cell can be tiled in such a way as to completely fill space.  For this reason, it is common to view periodic boundary conditions as a surrogate for the {\it bulk} or interior of a much larger model \cite{Adler59}. When boundary effects are of intrinsic interest, or when one wishes to model the behavior of confined systems, it is also possible to introduce additional terms for particle-boundary dynamics, e.g., reflecting boundary conditions that account for the behavior of particles bouncing off a wall \cite{Shen11,reflecting,boundarycondish}.  

Given a domain with boundary conditions, one next posits that $N$ particles with positions $\bq_i$ and momenta $\bp_i$ ($1\le i \le N$) occupy this space.  The collection of coordinates $\bbq = (\bq_1,...,\bq_N)$ and $\bbp = (\bp_1,...,\bp_N)$, which we also write as $\bx=(\bq_1,...,\bq_N,\bp_1,...,\bp_N)=(\bbq,\bbp)$, is known as the {\it phase space} of the system.  Typically, each $\bq_i$ and $\bp_i$ are 3-dimensional vectors, so that $\bx$ is a $6N$-dimensional vector.

From this vantage point, elementary counting arguments anticipate the first serious limitation of MD, namely its inability to model bulk systems.  Consider, for example, that  $48N$ bytes are required to store $6N$ coordinates in double precision ($8$ bytes per coordinate).  Thus, a volume $V\approx 1$ mm${}^3$ of carbon, amounting to $10^{20}$ particles, would consume roughly 5 billion terabytes of memory to store a single microstate! The situation does not meaningfully improve by scaling; for example $V=1$  $\mu$m${}^3$ and $N=10^{11}$, would still require 5 terabytes per microstate, to say nothing of the fact that many such microstates are needed for statistical averaging via Eq.~\eqref{eq:MCapprox}. Storage arguments aside, computational resources must also account for pairwise interactions, whose numbers scale quadratically with $N$.  Given these issues, it is clear that no computers can handle the processing requirements needed to model bulk systems in terms of their atomic constituents.

It follows from such considerations that practical MD simulations are limited to systems with $\mathcal O(10^6)$ atoms or fewer, corresponding to physical volumes on the order of cubic nanometers \cite{Ale12}.  This calls into question the validity of such simulations on two counts.  First, interatomic (e.g.\ electrostatic) forces can be appreciable at nanometer scales, implying that typical simulations may omit collective effects that occur over larger distances.  Second, the use of periodic boundary conditions allows for particles to interact with their periodic images (either directly or indirectly), thereby correlating interactions in a way that may be nonphysical and/or undesirable. In either case, the end result is that the system size may alter relative strength of interactions in play, thereby motivating assessment of $N$-dependent, ``finite-size'' effects. We will not discuss such issues in this review, instead referring interested readers to Refs.\ \ocite{Patrone16,SalI,SalII,Dario14}.

\subsection*{Inter-atomic Potentials}

As with all problems in dynamics, specification of forces is of prime importance. In MD systems, these are given in terms of the potential energy as a function of particle coordinates, $U(\bbq)$. The force on the $i$-th particle is then computed as the gradient with respect to $\bq_i$
\begin{align}
\bF_{\bq_i}(\bbq) = -\nabla_{\bq_i}U(\bbq) \label{eq:force}
\end{align}
The inter-atomic potential, $U(\bbq)$, encodes all information about the physics and chemistry of the constituent particles, as well as any external constraints on the system.  As such, it is generally computed not as a single function, but rather a sum over different types of interactions.  Schematically this may be expressed as
\begin{align}
 U(\bbq) = \sum_{i} U_1(\bq_i) 
   + \sum_{i<j} U_2(\bq_i,\bq_j)
   + \sum_{i<j<k} U_3(\bq_i,\bq_j,\bq_k) + \cdots \label{eq:genericff}
\end{align}
indicating one-body, two-body, three-body interactions, and so on \cite{allen89}.  Interactions with external fields, such as a constant electric field, are represented by the first term involving the sum over $U_1(\bq_i)$. We do not address these here. Below we briefly describe some of the more common 2- and 3-body potentials.  

\begin{figure}
\includegraphics[width=12cm]{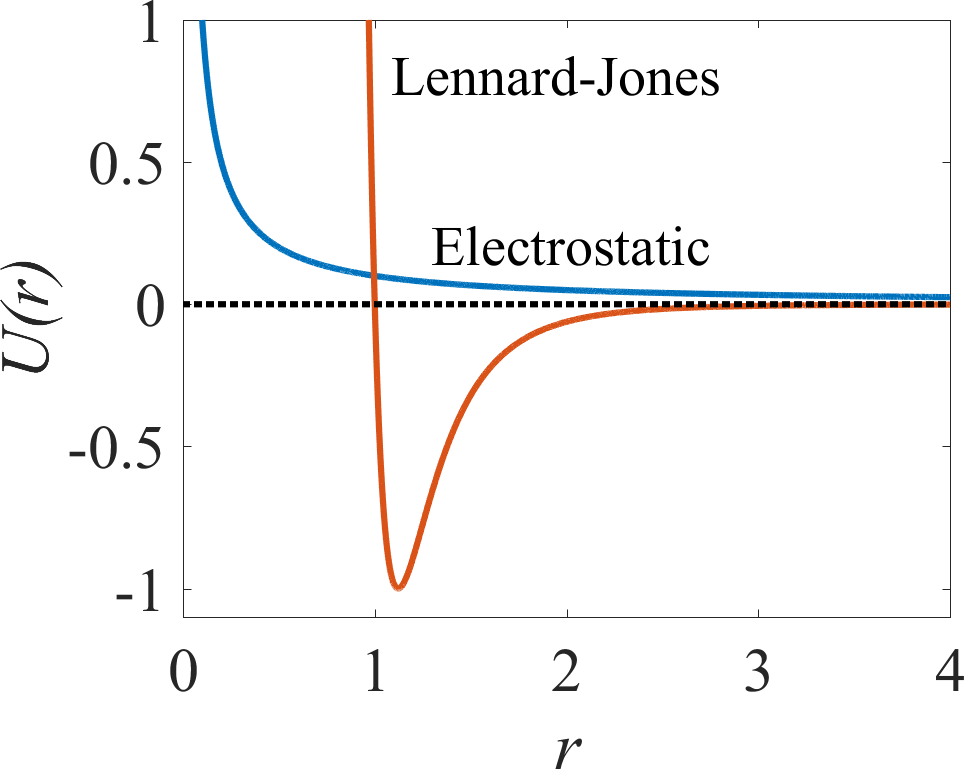}\caption{Illustration of Lennard-Jones (LJ) and electrostatic potentials.  For the former, we have set $\sigma=\epsilon=1$.  For the latter, we set $C\alpha_i\,\alpha_j=0.1$.  See Eqs.~\eqref{eq:electrostatic} and \eqref{eq:LJ}.  Note that the electrostatic potential has a ``softer'' divergence than the LJ potential as $r\to 0$.  The latter therefore better approximates hard-sphere interactions, which inhibit particles from coming closer than twice the particle radius. }\label{fig:lj}
\end{figure}

Common non-bonded, two-body interactions include electrostatic and van der Waals forces. For charged particles, the former is described by the familiar Coulomb potential
\begin{align}
U(\bq_i,\bq_j) = C\frac{\alpha_i\,\alpha_j}{r_{ij}} \label{eq:electrostatic}
\end{align}
where $\alpha_i,\ \alpha_j$ are net charges and $r_{ij}:=\|\bq_i-\bq_j\|$ is the distance between the atoms \cite{allen89,genMDandFF}. The proportionality constant, $C$, depends on choices of units in implementation. For otherwise neutral atoms and molecules, van der Waals interactions describe forces that arise from the spontaneous polarization of electron distributions  \cite{sakurai2011modern}. The {\em ab initio} evaluation of this potential requires a quantum mechanical computation solving the Schrodinger equation for the outermost electronic orbitals. This is rarely done. Instead, recognizing that the interatomic force can be both attractive and repulsive, an empirical model is used with a small number of parameters to be calibrated (or ``tuned'') to approximate the true forces for the given combination of atoms or molecules.  A favorite model of this type is the 6-12 Lennard-Jones potential, which takes the form \cite{genMDandFF}
\begin{align}
U(\bq_i,\bq_j) = 4\epsilon\left(
  \left(\frac{\sigma}{r_{ij}}\right)^{12}-\left(\frac{\sigma}{r_{ij}}\right)^6\right) \label{eq:LJ}
\end{align}
The $r_{ij}^{-6}$ decay at large distances is consistent with dipole interactions of quantum mechanics \cite{sakurai2011modern}, whereas the $r_{ij}^{-12}$ divergence for small separation approximates a hard-repulsive force; see Fig.~\ref{fig:lj}.  The location of potential minimum is determined by $\sigma$ and its depth by $\epsilon$.  These parameters are set to reflect the characteristic length and time scales for the system under consideration.  It is worth mentioning in passing that, while the 6-12 Lennard-Jones model is one of the most common in the MD community, its structure is a combination of heuristics and convenience.  We recommend reading Refs.\  \ocite{genMDandFF,allen89,binder95} for more details and alternatives.

Accounting for bonds between atoms introduces much more complexity into the potential model, requiring additional 2- and 3-body terms. In turn, each of these contributions are often modeled as a low-order Taylor series expansion about a nominal equilibrium point. For example,
\begin{subequations}
\begin{align}
U_b(\bq_i,\bq_j) &= K_2(r_{i,j}-r_0)^2+K_3(r_{i,j}-r_0)^3+K_4(r_{i,j}-r_0)^4 \\
U_a(\bq_i,\bq_j, \bq_k) &= C_2(\theta_{i,j,k}-\theta_0)^2+C_3(\theta_{i,j,k}-\theta_0)^3+C_4(\theta_{i,j,k}-\theta_0)^4 \\
U_{bb}(\bq_i,\bq_j, \bq_k) &= M(r_{i,j}-r_1)(r_{j,k}-r_2)
\end{align}
\end{subequations}
represent contributions from: bond-stretching, bond-angles, and bond-bond interactions. Deciding which terms to include requires deeper understanding from physical chemistry, taking into account the effects of the local environment within the molecule. Furthermore, the force-field parameters must be determined through independent calibration experiments involving comparisons of energy evaluations with {\em ab initio} potential energies, and comparison of full MD integrations with measured properties such as lattice constants. The expressions above are a subset of those used by the COMPASS (condensed-phase optimized molecular potentials for atomistic simulation studies) force-field first described in Ref.\ \ocite{sun:1998}.   Other popular force-fields include the Assisted Model Building and Energy Refinement (AMBER) used often for biomolecular simulations, and the Chemistry at Harvard Macromolecular Mechanics (CHARMM) set of force-fields tuned for protein dynamics.

Given the variety of methods for constructing $U(\bbq)$, it may be anticipated that the task of choosing a force field raises several questions related to UQ. For one, specific functional forms for $U(\bbq)$ may introduce undesirable effects due to limited structure, such as the single energy minimum or a fixed power-law repulsion explicit in Eq.~\eqref{eq:LJ}.  Moreover, the calibration process is generically imperfect, resulting in ``optimal'' force fields whose parameters nonetheless carry uncertainties.  In principal, these parametric uncertainties should be propagated into simulated predictions.  Finally, force fields are often used outside their domain of calibration, and it is rarely clear how to {\it objectively} define a most-appropriate potential in such situations.  Recently, strategies for addressing some of these issues have begun to emerge, especially in the context of coarse-grained MD \cite{cailliez:2011, Rizzi1,Rizzi2,Rizzi3,Rizzi4,PatroneRosch16,Louis,Carbone,Ghosh,Potestio,Faller2013,Brini,Farah,water1,Wu15,Angel13,Patra203}. In the following we assume that the force-fields are both fixed and suitable as given.

\subsection*{Hamilton's Equations}
\label{subsec:HamEquations}
With particles and forces in hand, Newtonian physics dictates how the system transitions between points in phase-space.  Specifically, one defines a {\it Hamiltonian} (or energy function) 
\begin{align}
H =\frac{1}{2} \sum_{i} \frac{\|\bp_i\|^2}{m_i} +U(\bbq) \label{eq:Hamiltonian}
\end{align}
where $U(\bbq)$ is the potential energy of the system, $m_i$ is the mass of the $i$th particle, and the Euclidean norm $\|\bp_i\|^2/(2 m_i)$ is the kinetic energy of the $i$th particle \cite{goldstein}. As discussed in the previous section, $U(\bbq)$ determines the forces acting on the particles via the gradient \eqref{eq:force}. These in turn determine the system evolution via Hamilton's formulation of Newton's equations (recalling that $\bx=(\bbq,\bbp)$),
\begin{align}
\dot \bx = \bj \nabla_{\bx}H \label{eq:HamiltonNewton}
\end{align}
where $\bj$ is a $6N \times 6N$ dimensional matrix defined as
\begin{align}
\bj = \begin{bmatrix}
\bzero & \b1 \\ -\b1 & \bzero
\end{bmatrix}
\end{align}
and $\boldsymbol 1$ is the $3N$ dimensional identity matrix. It is straightforward to show that Eq.~\eqref{eq:HamiltonNewton} reproduces the familiar equations
\begin{subequations}
\begin{align}
\dot \bq_i &= \bp_i / m_i \\
\dot \bp_i &= \bF_{\bq_i}
\end{align}
\end{subequations}

As a class of differential equations, Eq.~\eqref{eq:HamiltonNewton} displays a remarkably general property of energy conservation \cite{goldstein,Limecooler}, which will be important for our UQ discussions later.  In particular, 
\begin{subequations}
\begin{align}
\frac{\d}{\dt} H(\bbq(t),\bbp(t)) 
  & = \gradH \cdot \dotQP \\
  & = \gradH \cdot \bj \gradH \\
  & = -\gradHP\cdot \gradHQ + \gradHQ\cdot \gradHP \\
  & = 0
\end{align}
\end{subequations}
This observation has far-reaching consequences. For example, if the potential in Eq.\ \eqref{eq:Hamiltonian} is bounded from below,  $U(\bbq)\ge U_{\mathit{min}}$, then the solution to Eq. \eqref{eq:HamiltonNewton} exists for all time.  To prove this, assume otherwise, i.e., that some coordinate $\bq_i(t)\to \infty$ as $t \to T$.  It follows that $\bp_i=\dot \bq_i/m_i$ is also divergent and, so too, the kinetic energy.  However this cannot be so as the potential energy is bounded from below and the sum of both, kinetic and potential, is the constant (finite) Hamiltonian.  Such global control of a system of  ordinary differential equations (ODEs) is surprising in the face of the large dimension and non-linearity.

This grand statement aside, it turns out that Eq.~\eqref{eq:HamiltonNewton} resists further closed-form analysis. This is to be expected. Molecular dynamics may be viewed as a generalized version of the classical $N$-body problem, with the differences being that $N$ is extremely large and the inter-particle energies are more complicated than the gravitational $1/r$ potential. The point in making such a connection is that, as is well-known, this classical problem admits chaotic solutions for $N \ge 3$. More precisely, orbits with nearby initial conditions can diverge from one another at an exponential rate. This situation does not improve as $N \gg 3$ and the interactions become more complex. At an atomistic level, exponential divergence of paths for a single-particle billiard system was first proved by Sinai in 1970 \cite{sinai:1970}. Numerical experiments on Lennard-Jones systems performed soon after exhibited similar behavior \cite{stoddard:1973}. Thus, it is now taken as a given that molecular dynamics systems are chaotic, and likewise exhibit this sensitive dependence.

As analytical solutions to Eq.~\eqref{eq:HamiltonNewton} are generally impossible to find when $N > 3$, MD makes use of numerical integrators to advance time by discrete increments $h$, where the approximate solution at the $n$-th timestep is denoted $\bx_n\approx\bx(nh)$.  While a complete review of such algorithms is beyond the scope of this work, we find it convenient to consider two popular routines, the explicit fourth-order Runge-Kutta (RK4) method and the velocity Verlet (VV) algorithm. We introduce both here, and present a more detailed analysis in later sections.

Defining $\bF(\bx)=\bj \nabla_{\bx}H(\bx)$ for the moment, the 4-th order Runge-Kutta routine advances one full time step $h$ via a sequence of intermediate calculations
\begin{subequations}
\begin{align}
\bk_1 &= \bF(\bx_n) \label{eq:RKa}\\
\bk_2 &= \bF(\bx_n+\frac{h}{2}\bk_1) \\ 
\bk_3 &= \bF(\bx_n+\frac{h}{2}\bk_2) \\
\bk_4 &= \bF(\bx_n+h \bk_3) \\
\bx_{n+1} &= \bx_n+\frac{h}{6}(\bk_1+2\bk_2+2\bk_3+\bk_4) \label{eq:RKb}
\end{align}
\end{subequations}
The algorithm is classical and serves as a general-purpose workhorse for ODE integration \cite{iserles2009}. 

In comparison, the Velocity-Verlet algorithm advances momenta and coordinates according to 
\begin{subequations}
\begin{align}
\bbp_{n+\frac12} &= \bbp_{n}-\frac{h}{2}\nabla_{\bbq}U(\bbq_n) \label{eq:VV1}\\
\bbq_{n+1} &= \bbq_{n}+h \massM^{-1}\bbp_{n+\frac12} \label{eq:VV2}\\
\bbp_{n+1} &= \bbp_{n+\frac12} -\frac{h}{2}\nabla_{\bbq}U(\bbq_{n+1}) \label{eq:VV3}
\end{align}
\end{subequations}
where $\massM$ is the diagonal matrix of particle masses and we have reverted to the original coordinates ($\bbq$ and $\bbp$) \cite{VV}.  Informally, Eqs.~\eqref{eq:VV1}--\eqref{eq:VV3} may be viewed as: (i) keeping particle locations fixed and applying a constant force over a half time step to change momentum; (ii) allowing particles to drift for a full time step using the momentum computed in the first step; and (iii) update the forces and apply them to the momentum for another half step.  The intermediate variable $\bbp_{n+\frac12}$ is written for instructive purposes only and may be removed by substituting Eq.~\eqref{eq:VV1} into Eqs.~\eqref{eq:VV2} and \eqref{eq:VV3}, yielding $\bbq_{n+1}$ and $\bbp_{n+1}$ strictly in terms of their previous values.

The use of approximate methods such as Eqs.~\eqref{eq:RKa}--\eqref{eq:RKb} and Eqs.~\eqref{eq:VV1}--\eqref{eq:VV3} raises several questions. In the context of numerical analysis, it is reasonable to consider the extent to which a discrete trajectory $\{\bx_1,\ldots,\bx_n,\ldots\}$ agrees with its exact counterpart $\{\bx(t_1),\ldots,\bx(t_n),\ldots\}$ for each time step.  This issue is often addressed through the use of a {\it local error analysis} that quantifies the difference between  discrete and exact solutions when advanced by one time step from the same initial conditions.  Assuming generic initial conditions $\bx_0$, let $\Dh(\bx_0)$ be the result of a discrete time step of size $h$ starting from $\bx_0$, and denote by $\Fh(\bx_0)$ the exact flow from the same starting point. Then numerical analysis seeks to control errors via 
\begin{align}
|| \Dh(\bx_0)-\Fh(\bx_0)|| < Ch^{r+1}\label{eq:localerror}
\end{align}
Here $C$ is a constant, $r$ is the order of the algorithm, and the bound is to hold independent of $\bx_0$.  From Eq.\ \eqref{eq:localerror}, one can arrive at the desired comparison between $\bx_n$ and $\bx(nh)$ by considering accumulation of error \cite{iserles2009,Limecooler}, which is bounded by $\mathcal O(h^r)$ over long timescales. Such statements are staples of upper level numerical analysis courses and motivate examination of higher-order methods so as to increase $r$. As we will see, however, this approach is typically viewed as misguided in the context of MD.

Juxtaposing a local error analysis is the more statistical idea that MD needs to preserve the structure of Hamilton's equations in such a way that the discrete solution explores approximately the same region of phase space as $\bx(t)$.  In other words, correspondence of the form $\bx_n\approx\bx(nh)$ is less important so long as both provide comparable sets of points $\{\bx_i\}$ for purposes of evaluating ensemble averages as per Eq.~\eqref{eq:MCapprox}.

The technical machinery underlying this requirement is of a decidely different character. To motivate it we consider $\by=\phi(\bx)$ to be an invertible and differentiable transformation from $\bx$ to a new phase-space point $\by$.  Through repeated application of the chain rule one may show that the trajectory viewed in coordinates $\by$ will obey the transformed equation
\begin{subequations}
\begin{align}
\dot \by 
   &= (\nabla_{\bx}\phi) \dot \bx\\
   &= (\nabla_{\bx}\phi) \bj \nabla_{\bx} H(\bx)\\
   &= (\nabla_{\bx}\phi) \bj (\nabla_{\bx}\phi)^{\rm T} \nabla_{\by}\tilde{H}(\by)
\end{align}
\end{subequations}
Here the Hamiltonian in terms of $\by$ is given by $\tilde{H}(\by) = \tilde{H}(\phi(\bx))=H(\bx)$, and the Jacobian is defined by $(\nabla_{\bx}\phi)_{i,j} = \partial \phi_i/\partial x_j$. The crucial observation is, if the Jacobian satistfies$(\nabla_{\bx}\phi) \bj (\nabla_{\bx}\phi)^{\rm T} = \bj$, then the transformed ODE in the $\by$ coordinates is again a Hamiltonian system.  For mathematical reasons it is more common to write the transpose on the other side and define a transformation $\phi(\bx)$ to be {\it symplectic} if it satisfies
\begin{equation}
(\nabla_{\bx}\phi)^{\rm T} \bj (\nabla_{\bx}\phi) = \bj \label{eq:symplectic}
\end{equation}
One motivation for identifying this class of transformations is that, as we have seen, they preserve the Hamiltonian character of the ODE system. At a more fundamental level, focusing attention on symplectic transformations ushers geometrical reasoning into the analysis of Hamiltonian systems.

Returning to the problem of trajectory comparison, it turns out that the continuous flow map of a Hamiltonian system, $\Fh(\bx)$, is symplectic. While confirmation requires verifying the identity Eq.\ \eqref{eq:symplectic}, intuitively one anticipates it to be so as evolving a Hamiltonian system forward in time does not change its Hamiltonian structure. This fact leads one to analyze discrete flow-maps entailed by ODE integration algorithms, $\Dh(\bx)$, paying attention to their symplectic character. Such investigations gave rise to a branch of analysis by the name of geometric numerical integration that first took hold in the 1980's and continues to be active today \cite{hairer:2003}. While further consideration would take our discussion too far astray, we do draw attention to the key result in the field which exists in the form of a theorem which we state informally as: 

\begin{thm}[Benettin and Giorgilli, 1994] Assume we are given, $\Dh,$ a smooth transformation near the identity. If $\Dh$ is symplectic, then there exists a modified Hamiltonian $\tilde{H}$ such that iterates of $\Dh$ agree with the exact flow for $\tilde{H}$ for exponentially long times. In other words, for $h$ sufficiently small there exists $\mu$, $\gamma$, and $\epsilon$ such that
\begin{align}
\|\Dh^{^k}(\bx_0)-\tilde{{\cal F}}_{kh}(\bx_0)\| < C \exp(k\mu\epsilon)\exp(-\gamma/\epsilon)
\end{align}
\label{thm:shadowH}
\end{thm}

We refer to the original reference for the precise statement and proof \cite{benettin:1994}. To understand the importance of this result in the context of MD one reasons as follows. Given the molecular system, use a symplectic integration algorithm to advance time.  For small enough $h$, each time step will result in a small change in phase space.  In other words, the time-stepping algorithm is a symplectic transformation near the identity. From Theorem \ref{thm:shadowH}, it follows that there exists a modified Hamiltonian such that the resulting collection of discrete iterates, $\{\bx_1,\bx_2,\ldots\}$, are the exact trajectory of this modified system. This remarkable theorem is one of the deeper facts in this business and requires time to digest.  Despite its implications, however, many questions remain unresolved; e.g.\ is the phase-space of the modified Hamiltonian ``close'' in some sense to its unmodified counterpart? In the tutorial section, we explore these issues in the context of the harmonic oscillator, where the abstract flow-maps $\Fh$ and $\Dh$ become matrices and the analysis can be carried out explicitly in full detail.

From a UQ perspective, the point of highlighting these issues is to emphasize that the MD community is at odds with more traditional numerical analysis that attempts to quantify the accuracy of particle trajectories {\it per se}.  In particular, community practice has shown that in many cases, it is perhaps more important to assess the consistencies between the discrete flow and the target ensembles as predicted by statistical mechanics.  This motivates us to consider these ensembles more carefully below, both from a theoretical and UQ perspective in the subsequent sections and tutorials.

In passing, we mention the central role of the time-step $h$ in this discussion. It is reasonable that a discrete trajectory should adequately sample the fastest motions if it is to be a good approximation of its continuous counterpart. Thus, the timescales of the physics in question set a limit of reasonable choices of $h$. In many systems, the fastest motions correspond to oscillations of C-H or N-H bonds, which have frequencies on the order of $10^{14}$ Hz \cite{Schlick98}.  Thus, $h=1$ fs (or $1/h=10^{15}$ Hz) is an approximate upper bound on admissible $h$ that still sample the relevant physics.  Unfortunately, however, this timestep is so small that current high-performance computers can only simulate systems over the course of nanoseconds to microseconds, i.e.\ $\mathcal O(10^{-9})$ seconds to $\mathcal O(10^{-6})$ seconds.  Thus, attempts to model experimental procedures such as annealing or mechanical property tests subject simulated materials to astronomical cooling and loading rates.  Quantifying uncertainties associated with these time-scale limitations remains an open problem in many cases.

\subsection*{Thermodynamic Ensembles}
\label{subsubsec:ensemble}

It is a fundamental result of equilibrium statistical mechanics that $\pdf(\bx)$ is a function of the Hamiltonian $H(\bx)$.  The precise form of the relationship between $H$ and $\pdf$ depends on the extent to which the system is coupled to an environment.  Three cases are commonly considered \cite{landau,Pathria}.  

\begin{figure}
\includegraphics[width=16cm]{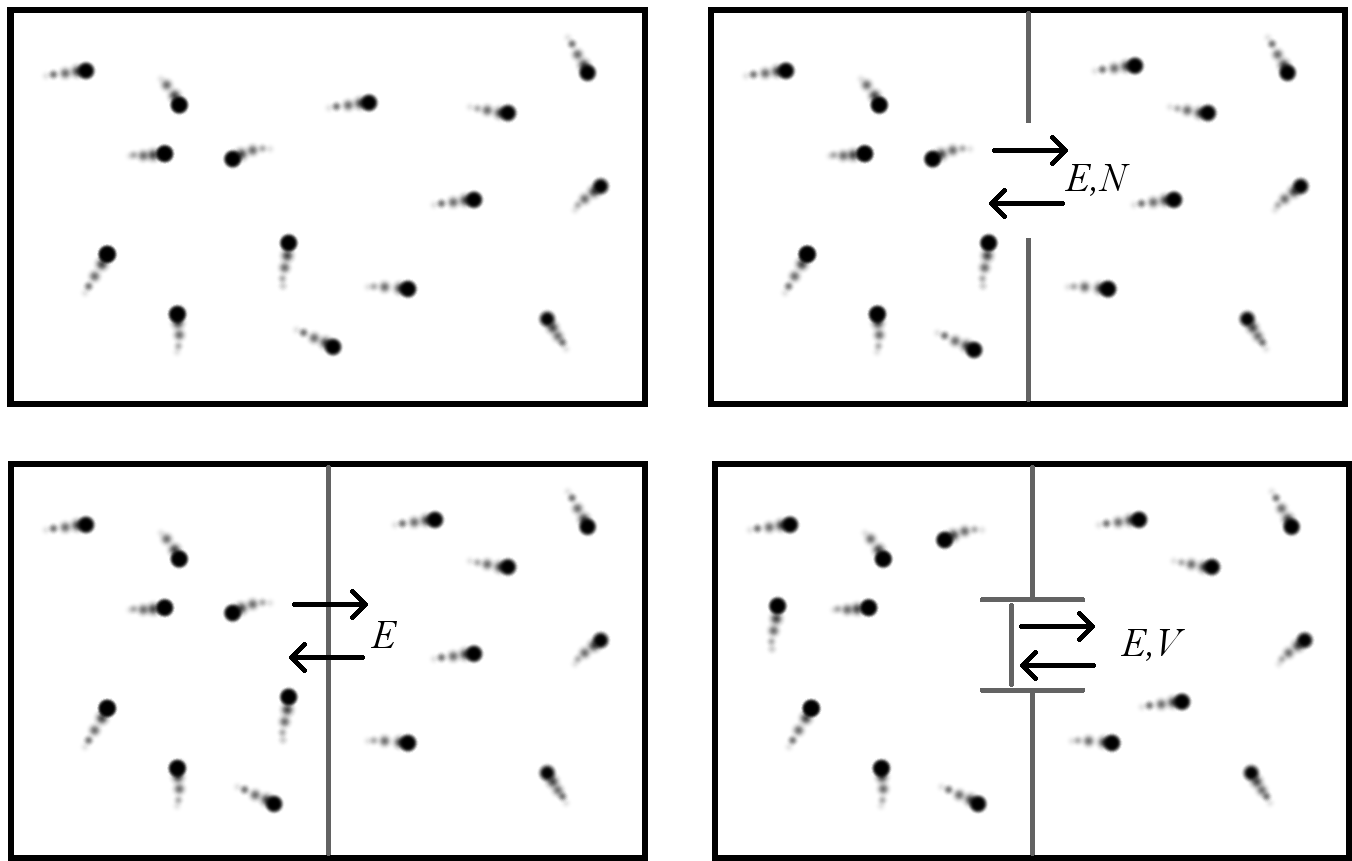}\caption{Schematic of three thermodynamic ensembles.  Top Left: The microcanonical ensemble describes an isolated system in a closed container.  Bottom Left: The canonical ensemble describes a system that can exchange energy (but not particles) with an external heat bath or reservoir.  Right: The grand canonical ensemble describes a system that can exchange both energy and particles with a reservoir (top).  In MD, it is more common, however, to model a system whose volume (as opposed to number of particles) can change (bottom).} \label{fig:ensembles}
\end{figure}

In the {\it microcanonical} (or NVE) ensemble, the number of particles $N$, volume $V$, and energy $E$ are fixed quantities, corresponding to a system that is isolated from the world.  In this situation, the Hamiltonian $H=E$ is constant, and all microstates $\bx$ satisfying $H(\bx)=E$ have equal probability.  From the perspective of MD, this is the easiest ensemble to simulate, since it only entails integrating Eq.~\eqref{eq:HamiltonNewton} without any additional constraints or assumptions about the system\blfootnote{The careful reader will realize that we have glossed over two key issues, namely conservation of linear and angular momentum.  In statistical mechanics, it is convenient to discard these principles by defining the system relative to a fixed container whose walls are assumed large (relative to the system size) and can thus act as a source/sink for momentum \cite{landau}.  In MD, the situation is slightly more tenuous when periodic boundary conditions are used.  In particular, linear momentum is conserved, but angular momentum is not \cite{Limecooler}.  This latter violation is often simply accepted as unavoidable.}.  See Fig.~\ref{fig:ensembles}.

If the system is allowed to exchange energy, but not particles or volume, with a reservoir at some fixed temperature $T$, the corresponding system is in the {\it canonical} (or NVT) ensemble; see Fig.~\ref{fig:ensembles}.  Statistical mechanics establishes that the probability of the system being in any given microstate is given by
\begin{align}
\pdf_{NVT}(\bx) =\frac{1}{\mathcal Z} \exp\left[\frac{-H(\bx)}{k_BT} \right] \label{eq:boltzmannfactor}
\end{align}
where $k_B$ is the Boltzmann constant and the normalization constant $\mathcal Z$ is known as the partition function.  Although not of intrinsic value to our UQ discussion, $\mathcal Z$ is an important quantity in its own right and is related to many thermodynamic properties \cite{landau,Pathria}.

If the system is allowed to exchange both particles and energy with a reservoir, then it is a {\it Grand Canonical} ensemble; see Fig.~\ref{fig:ensembles}.  For practical purposes in simulations, it is often more convenient to fix $N$ and the pressure $P$ and instead allow the volume to fluctuate; such simulations are denoted NPT \cite{Andersen80,Parrinello81}.  Again, it is a fundamental result of statistical mechanics that the corresponding probability is defined via
\begin{align}
\pdf_{NPT}(\bx) \propto \exp \left[\frac{-H(\bx)-PV}{k_BT}  \right]\label{eq:nptprob}
\end{align}

Given that the NVT and NPT ensembles assume the existence of an external (and much larger) reservoir, the finite-size limitation of MD raises questions of how to generate trajectories consistent with Eqs.~\eqref{eq:boltzmannfactor} and \eqref{eq:nptprob}.  Directly modeling the reservoir is intractable, since this would amount to running a much larger simulation than the system of interest.  Rather, the standard approach involves augmenting the system description  with additional degrees of freedom that act as surrogates for the reservoir.  In the literature, these degrees of freedom are often referred to as {\it thermostats} and {\it barostats}, which control the temperature and pressure respectively \cite{Andersen80,Parrinello81,Nose,Hoover,Andersen2,schlick2010molecular}.  

While it is beyond the scope of this chapter to describe every such algorithm in detail, we find it useful to consider the Nos{\' e} thermostat \cite{Nose}, variations of which are widely used throughout the community (see Ref.\ \ocite{Limecooler} for a more comprehensive overview).  The central idea is to introduce an additional degree of freedom $s$, which acts like a ``virtual'' particle with a momentum $p_s$ and mass $\mathcal Q$.  Then, by augmenting and rescaling the dynamical variables according to $\tilde \bx = (\bbq,\bbp',s,p_s)$, where
\begin{subequations}
\begin{align}
\bbp &= \bbp'/s \\
\dt &= \dt'/s
\end{align}
\end{subequations}
the Hamiltonian is reformulated as
\begin{align}
H_{N} =\frac{1}{2} \sum_{i} \frac{\|{\bp'}_i\|^2}{s^2 m_i} +U(\bbq) + \frac{p_s^2}{2\mathcal Q} + (3N+1) k_BT\ln(s) \label{eq:NHHamiltonian}
\end{align} 
Nos{\' e} was able to show that by evolving the extended system in the microcanonical ensemble, the projected probability  $\pdf(\bx)$ (distinguish from $\pdf(\tilde \bx)$, which is constant) is a canonical distribution {\it under the assumption that the simulation is ergodic} \cite{Nose}.

Informally, the mechanism behind the Nos{\' e} thermostat can be understood by examining the corresponding equation for $\dot p_s$, which reads
\begin{align}
\dot p_s = \frac{2}{s} \left[ \sum_i \frac{||\bp_i||^2}{2 m_i} - \frac{3N+1}{2}k_BT \right] = \frac{2}{s} \left[ \sum_i \frac{||{\bp'}_i||^2}{2 m_is^2} - \frac{3N+1}{2}k_BT \right] \label{eq:psdot}
\end{align}
in terms of both the original and rescaled variables.  In statistical mechanics, the principle of equipartition of energy states that at temperature $T$, the average energy in any given degree of freedom (i.e.\ associated with translational motion, rotational motion, etc.) is $(1/2)k_BT$.  In other words, thermal energy is, on average, distributed evenly across all the momentum modes that can store it.  Thus, the average kinetic energy of the $3N+1$ momentum degrees of freedom (including $p_s$) should be $\frac{3N+1}{2}k_BT$, i.e.\ the term on the far right side of Eq.~\eqref{eq:psdot}.  If the kinetic energy $\frac{1}{2}\sum_i \frac{\|\bp_i\|^2}{m_i}$ is greater than its thermal average, the value of $p_s$ increases, which eventually decreases the kinetic energy via the rescaling coefficient $\frac{1}{s^2}$.  Similarly, low values of kinetic energy decrease $s$, thereby increasing the former.  In this way, the Nos{\' e} thermostat is a negative feedback control that attempts keep the thermal kinetic energy from deviating too much from its mean value predicted by thermodynamics.

Importantly, this discussion reveals several key assumptions underlying the thermostat.  For one, the temperature control is tied directly to the {\it kinetic energy alone}, a feature that is common to many thermostats \cite{Nose,Hoover,Andersen2,Berendsen}.  Thus, where ergodicity is required but not proved (e.g.\ as in the Nos{\' e} thermostat), we are not guaranteed that $\pdf(\bx)$ actually samples a canonical distribution.  It is known, for example, that a one-dimensional harmonic oscillator may still exhibit periodic orbits in phase space when coupled to a Nos{\' e} thermostat \cite{Noseoscillator}.  Stated differently, thermostats based on controlling the kinetic energy may not satisfy the more general equipartition result \cite{Limecooler}
\begin{align}
\Big\langle x_m \frac{\partial H}{\partial x_n} \Big\rangle = \delta_{m,n} k_BT \label{eq:genequipartition}
\end{align}
Moreover, thermostats invariably introduce coupling constants (e.g.\ $\mathcal Q$ in Eq.~\eqref{eq:NHHamiltonian}) that affect the rate at which heat is added or removed to the system.  The values for these couplings need to be considered carefully when running simulations in order to ensure that they do not lead to nonphysical responses to the thermostat; see Refs.\ \ocite{Limecooler,Icecube,Icecube2,Mor08} for discussion of these issues.

\subsection*{Where does this leave us?}
In providing an overview of MD, the reader has hopefully come to appreciate both its ingenuity and the sometimes bold assumptions that underpin its implementation. In view of this, it should be clear that a fundamental problem for UQ of MD is the breadth of issues that must be addressed for a full treatment thereof.  Eventually, this extent is so great that we are all forced to accept our lack of knowledge and proceed nonetheless.  Thus, we are reminded of the following quote regarding MD \cite{frenkel2001}: ``... there is clearly still a corpse in the closet. We believe this corpse will not haunt us, and we quickly close the closet.''  We wish to add to this, however, that the closet door should be opened from time to time, skeletons dusted off and re-examined for those that are no longer scary.  Through our work, we hope to show that with progress, some will even be removed from the closet, so that eventually none will remain.

\section*{Uncertainty Quantification}
\label{sec:UQsec}
Having provided a high-level overview of MD, we now turn our attention to the theory of uncertainty quantification.  While UQ is a large and developed field in-and-of itself, applications to MD are still relatively new.  As a result, much of the existing UQ literature is devoted to topics that may not be of intrinsic interest to computational material scientists.  Thus, our primary goal in this section is to familiarize the reader with ideas that will be useful in the tutorials, and more generally, for applications to MD.  For a good overview of UQ, we refer the reader to the recent book by Smith \cite{smith2013}. 

The discussion that follows is divided into two subsections.  In the first, we provide a general overview of UQ, with the purpose of highlighting the various perspectives from which scientists have addressed this topic.  In the second, we review key concepts from probability theory and statistics in order motivate inference techniques, which play a large role in the tutorials.  We also discuss one approach for uncertainty propagation.

\subsection*{What is UQ?}
\label{subsec:whatisUQanways?}

In beginning the discussion, it is natural to ask, ``What is uncertainty quantification?''  Perhaps surprisingly, there is some disagreement within the community as to the best answer \cite{Pate96}.  For our purposes, we find it convenient to adopt a practical perspective: {\it uncertainty quantification is the task of assessing and building confidence in a result for the purposes of being able to make a decision.}  While this is perhaps an overly general definition, it provides a clear motivation of what one seeks to gain from the analyses, thereby allowing one to pick those tasks that are most informative.  It also reveals that UQ is, to a certain extent, a subjective endeavor.  Time and money dictate that one cannot indefinitely collect data and analyze results, so that the thoroughness of a UQ analysis depends on the belief---hopefully shared by modeler and decision maker---that the results are useful.

A recurring theme within the UQ community recognizes that uncertainties may be distinguished as originating from two principal sources. {\it Aleatoric uncertainty} is uncertainty due to a fundamentally random phenomenon \cite{Pate96,Hora96}. Elementary examples of this include flipping a coin or rolling dice. The underlying idea is that, while it is possible to identify different states of a phenomenon, which state is assumed by the system is determined only once the event occurs.  Furthermore and critically, repeating the experiment will precipitate different outcomes. In the present work,  fluctuations of measurable quantities are a representative example of aleatoric uncertainty. By contrast, {\it epistemic uncertainty} refers to uncertainty arising from incomplete knowledge \cite{Pate96,Hora96}. As a simple example, one might consider the height of a friend. While this quantity may be assumed to be fixed, its precise value is generally estimated as being within some range. This range may be taken as a measure of epistemic uncertainty. In contrast to aleatoric uncertainty, epistemic uncertainty can be reduced by acquiring more knowledge, e.g., in making careful measurements. Furthermore, it may be assumed that the quantity is fixed from one instance to the next. There is no need to posit hypothetical, multiple realizations for interpretation. In the context of MD, the force constants used to describe the interatomic potentials may be viewed as a source of epistemic uncertainty. 

While such high-level distinctions may seem academic, there are practical ramifications.  For one, common practice uses the tools of probability and its calculus to quantify both forms of uncertainty. While there is consensus that this is an appropriate mathematical structure for characterizing aleatoric uncertainty, some question its suitability for epistemic uncertainty. An elementary example illustrates this point. Consider, the case in which the length of an object is known to be less than one unit with no other information available. A plausible model for this situation represents this (epistemic) uncertain length as a uniform random variable over the interval $[0,1]$. A number of confounding issues arise. First, if the objects are manufactured by a consistent source, then one may question whether it is appropriate to model the joint distribution of multiple copies as independent draws from this probability law. Yet, many Monte Carlo sampling schemes do just this. One could argue that the problem is one of independence, and thus traditional probability calculus can accommodate this situation by accounting for correlations between uncertain quantities. However, there is an issue that lies even deeper still. Returning to the example, assume further that there are two manufacturing pathways such that two objects may be made independently of one another. Placing the objects side-by-side such that their lengths add, what is the best description of the uncertain outcome that would result? If the isolated objects are modeled as independent, uniform random variables on $[0,1]$, then standard arguments lead to the result that their sum be modeled as a symmetric triangular distribution on $[0,2]$. While the range of $[0,2]$ is indisputable, heuristic reasoning suggests that a uniform distribution on this interval is a more suitable description of the state of knowledge on the system. Stated differently, given such limited information from the outset, on what basis can one surmise that a combined length of 1 should be the most likely outcome? From the perspective of analysis, the crux of this issue lies in the fact that probability calculus requires that uncertainty of an combination of random variables is distributed such that its total mass is one. Sometimes, as is the case with the simple addition example above, this invariance requirement induces a structure that conflicts with other forms of heuristic reasoning. Some researchers have begun investigating what happens when conservation of total probability is not required. In more colorful moments the situation is compared to the suspension of the parallel line postulate which famously gave birth non-Euclidean geometry \cite{ferson}. While this pursuit is intriguing, we nevertheless follow common practice and describe all uncertainties using random variables and their calculus.

As a final point we note that the decision to model uncertainty as a random variable does not fully resolve all philosophical issues, but rather shifts the discussion to the nature of probability itself.  In particular, two perspectives are common throughout the UQ community. The {\it frequentist} interpretation of probability views randomness as an inherent property of the world that can be described by relative frequencies of possible outcomes. In the simplest case, if a random variable can take on only a finite set of values, then it is sensible to discuss the relative frequency that each occurs. More mathematically, let $x$ be one of the allowed discrete values and assume $N$ independent realizations $\{y_1,y_2,\ldots,y_N\}$ of the random variable.  Then the probability of event $y$ is defined by the limit
\begin{align}
\pdf(y) =\lim_{N\to \infty} \sum_{i=1}^N \frac{\delta_{y,y_i}}{N} \label{eq:freqprob}
\end{align}
where $\delta_{y,y_i}=1$ if $y=y_i$ and zero otherwise. From this operational definition it follows that $\pdf(y)$ is between $0$ and $1$, and that the sum of $P(y)$ over all possible values $\{y_1,\ldots,y_N\}$ is one. Technical mathematical machinery is required to extend this construction to the situation where the possible outcomes are countably infinite, and finally outcomes containing a continuum. Nevertheless, analysts were up to this task and much of its codification is attributed to Kolmogorov in the 1930's.  For our discussion the key point is that, for the frequentist, the probability of any outcome represents the relative frequency of that outcome that would arise from a large collection of independent realizations of the system of interest.

In contrast, the {\it Bayesian} interpretation of probability explicitly views randomness (and thus uncertainty) as an expression of our lack of knowledge about the system of interest \cite{gregory:2005}. For the Bayesian, probabilities $\pdf(x)$ of various events $x$ are not inherent properties of the events themselves, but rather a tacit admission that our understanding of the world is incomplete.  Thus, Bayesians often speak of updating beliefs, and thereby probabilities, as new information is gained \cite{kennedy,Bayesian1,Bayesian2,Bayesian3}.  The mathematical machinery for this is synonymous with conditional probabilities. In brief, if $A$ and $B$ are two events belonging to a joint probability space, then one can evaluate the probabilities $P(A\cap B)$, $P(A)$, and $P(B)$. Importantly, the conditional probabilities $P(A|B)$ and $P(B|A)$ are defined by the following equalities
\begin{align}
 P(A\cap B) = P(A|B)P(B) = P(B|A)P(A)
\end{align}
Thus, from an axiomatic perspective, the conditional probability is simply the ratio of the probability of the joint occurrence of two events to the probability of one event in isolation. Bayes' theorem isolates the second equality, writing it as
\begin{align}
P(A|B) = \frac{P(B|A)P(A)}{P(B)} \label{eq:Bayes}
\end{align}
and elevates the relationship to one of epistemology.  In other words, $P(A|B)$ is interpreted as a quantification of our belief that event $A$ will happen (or already has) given the occurrence of $B$.  Although not a tool we pursue here, Bayesians take this perspective further by proposing an iterative scheme on the basis Eq.~\eqref{eq:Bayes}, formalizing the notion that $P(B|A)$ is a likelihood which updates a prior probability $P(A)$ to yield $P(A|B)$ as a posterior distribution describing our improved state of knowledge, given some measurements of $B$.  

We conclude this overview with definitions of primary UQ concepts and terminology. In the 1990's, as the idea of UQ was expanding into new application domains, this terminology was subject to debate \cite{oberkampf:2010}. More recently a consensus has emerged. With some exceptions, the following definitions closely adhere to the those appearing in the standards guide created by the American Society of Mechanical Engineers (ASME) in 2006 \cite{ASME}:
\begin{itemize}
\item Verification: The process of determining that a computational model accurately represents the underlying mathematical model and its solution. 
\item Calibration: The process of adjusting parameters in a computational model to improve agreement with data obtained from an independent source--either experiment or a different computation.
\item Validation: The process of determining the degree to which a model is an accurate representation of the real world, given an intended use-case for the former. 
\item Uncertainty Quantification: The process of determining the uncertainty in a quantity of interest accounting for all sources of uncertainty upon which the quantity depends. 
\end{itemize}
In the context of MD, verification could encompass the following questions: Does my code have a programming bug? Is the simulated system representative of a chaotic, many-body system?  Is the simulation representative of a thermodynamic ensemble? Calibration is used to determine inputs such as parameters of the interatomic potential $U(\bbq)$ or the thermostat mass $\mathcal Q$ in Eq.~\eqref{eq:NHHamiltonian}.  Validation addresses questions such as: Does the simulation predict a given experimental result? What is the systematic error of the simulation relative to experiment? Here the comparison to experiment is critical.  Finally the ASME definition of ``uncertainty quantification'' above is more restrictive than the one we proposed at the outset and amounts, e.g., to the task of propagating uncertainties in model parameters to corresponding outputs.  Here we use the terms ``uncertainty propagation'' and ``inference'' to describe such tasks, reserving the phrase ``uncertainty quantification'' as an umbrella term that describes the overall process of building confidence in a prediction.

In general, the sequence of tasks---verification, calibration, validation---is a natural ordering of analyses that most modelers are at least informally aware of \cite{ASME}. Verification must be done first to confirm that the simulation is faithful to the conceptual model.  Calibration follows as an independent task, since the process of tuning free parameters during validation is questionable.  Nonetheless, neither our definitions nor their ordering are universal within the community, and the reader should be aware of this issue.  For example, the distinction between verification and validation may not always be clear; e.g.\ determining that a simulation represents a thermodynamic ensemble can be interpreted as validation, since it suggests comparison with a certain class of experiments.  Moreover, our definition of calibration differs from that of ASME insofar as the latter restricts comparison to experimental data alone.  In the context of MD, this precludes comparison with other simulations, which are often used as the basis for coarse-graining techniques and reduced-order modeling hierarchies \cite{kennedy,Calibration1,Calibration2,Calibration3,OilCal,PatroneRosch16,Kriging1,CalVal1}.  Nonetheless, these issues largely amount to semantics, and ultimately we take recourse to our general perspective: if a given analysis is useful for decision-making, then it is a valid type of UQ. In the next section, we turn to these more practical issues, introducing concepts needed to actually perform UQ analyses.

\subsection*{Tools for uncertainty quantification}
\label{subsec:tools}

\subsubsection*{Maximum likelihood estimation}

In statistics, {\it inference} is the task of deducing one or more parameters $\phi$ defining a probability density function (PDF) $\pdf(\bz|\phi)$ given $N$ realizations  $\bz_i$.  If $\bz_i$ is the output of a simulation, then the $\phi$ may be of intrinsic value in quantifying uncertainties, e.g.\ if the standard deviation $\sigma \in \phi$.  Moreover, inference is often used in uncertainty propagation when we wish to estimate uncertainty in some function $f(\bz)$. 

A common approach for inferring $\phi$ is the {\it maximum likelihood estimate} (or MLE), where one computes
\begin{align}
\phi^{\star} = \argmin_{\phi} \left\{- \log\left( \prod_{i=1}^N \pdf(\bz_i|\phi) \right) \right\} \label{eq:genMLE}
\end{align}
In this expression, $\mathcal L(\phi|\{z_i\}) =\prod_i \pdf(\bz_i|\phi)$ is called the likelihood function (or likelihood of $\phi$ given the $\{\bz_i\}$).  MLE requires a functional form for $\pdf(\bz|\phi)$ as an input, which reveals that Eq.~\eqref{eq:genMLE} is in effect a {\it statistical model} of $\bz$ subject to the limitations and assumptions underlying the choice of $\pdf(\bz|\phi)$.

In order to better illustrate how MLE works, we consider the case in which $\pdf(\bz|\phi)$ is a Gaussian PDF and take $\phi=(\mu,\sigma)$, i.e.\ the mean and standard deviation.  One finds
\begin{align}
\phi^{\star} = \argmin_{\phi} \left\{C + N\log(\sigma) + \frac{1}{2\sigma^2}\sum_{i=1}^N (\bz_i - \mu)^2 \right\} \label{eq:GaussianMLE}
\end{align}
where $C$ is a constant that does not depend on $\mu$ or $\sigma$.  Denoting $A=C + N\log(\sigma) + \frac{1}{2\sigma^2}\sum_{i=1}^N (\bz_i - \mu)^2$, it is straightforward to solve this minimization problem by computing $(\partial_{\mu},\partial_{\sigma})A = 0$ and solving the resulting equations.  We leave it to the reader as an exercise to show that
\begin{align}
\mu^{\star} &= \frac{1}{N}\sum_{i=1}^N \bz_i \label{eq:sampmean} \\
(\sigma^{\star})^2 &= \frac{1}{N}\sum_{i=1}^N (\bz_i - \mu^{\star})^2 \label{eq:sampvar}
\end{align}
That is, MLE returns the familiar {\it sample mean} and {\it sample variance}\blfootnote{Technically speaking, Eq.~\eqref{eq:sampvar} is slightly biased insofar as $\E[(\sigma^{\star})^2]> \sigma^2$.  The problem can be remedied by replacing the factor of $1/N$ with $1/(N-1)$ in Eq.~\eqref{eq:sampvar}.  We leave it to the reader to prove this.} as its estimates for $\sigma$ and $\mu$.

Closer inspection of Eq.~\eqref{eq:GaussianMLE} reveals a similarity with the {\it method of least-squares}.  In particular, we could posit that $\bz$ is in fact a {\it random function} of some other variable such as temperature $T$, so that our model becomes
\begin{align}
\bz(T) = \mu(T) + \sigma(T) \mathcal N(0,1) = \mathcal N(\mu(T),\sigma(T)^2)
\end{align}
where $\mathcal N(\mu,\sigma^2)$ is a Gaussian random variable with mean $\mu$ and variance $\sigma$.  Thus, if $\bz_i=\bz(T_i)$ are random realizations of $\bz$ at different temperatures $T_i$, we can specify mean and variance functions $\mu=\mu(T,\phi_1)$ and $\sigma^2(T,\phi_2)$ [where $\phi=(\phi_1,\phi_2)$], which yields the MLE equations
\begin{align}
\phi^{\star} = \argmin_{(\phi_1,\phi_2)} \left\{C + \frac{1}{2}\sum_{i=1}^N \log\big(\sigma(T_i,\phi_2)\big) + \left(\frac{\bz(T_i) - \mu(T_i,\phi_1)}{\sigma(T_i,\phi_2)}\right)^2 \right\} \label{eq:GaussianLS}
\end{align}
When $\sigma(T)=\sigma$ is constant, Eq.~\eqref{eq:GaussianLS} is simply a least-squares estimate for the parameters $\phi_1$ that define the function $\mu(T,\phi_1)$.  Note that the differences $\bz(T_i)-\mu(T_i,\phi_1)^{\star}$ are {\it residuals to the fit}.  In the following sections, we consider refinements and variations on these procedures.

We note that in general, Eqs.~\eqref{eq:genMLE} and \eqref{eq:GaussianLS} cannot be solved for generic choices of PDFs $\pdf(\bz)$ or mean and variance functions.  Nonetheless in typical applications, software packages such as Matlab and Octave provide optimization routines that can be used to solve the associated MLE problems.  Thus, we assume that such tasks are tractable.  

As a final aside, we also note that because $\mu^{\star}$ is a sum over random variables, it is itself a random quantity that can also deviate from the true mean $\mu$.  Intuitively, however, as the number of samples $N$ becomes large, we expect $\mu^{\star}$ to be an increasingly good approximation of $\mu$.  To quantify this, we can compute the variance $\epsilon^2=\Var[\mu^{\star} - \mu]$ over the probability density $\pdf(\bz)$.  We leave it to the reader to show that 
\begin{align}
\epsilon^2 = \sigma^2/N \label{eq:meanvar}
\end{align}   
namely, the variance of the error scales as $1/N$, which vanishes in the limit $N\to \infty$.  If we replace $\sigma^2$ with its sample variance according to Eq.~\eqref{eq:sampvar}, we arrive at the standard error
\begin{align}
\epsilon^2_s &= \frac{1}{N^2}\sum_{i=1}^N (\bz_i - \mu^{\star})^2 \label{eq:standerr}
\end{align}
which estimates the amount to which $\mu^{\star}$ deviates from the true mean.  

\subsubsection*{Spectral approach to non-parametric inference}
\label{app:smc}

In certain UQ applications, one requires precise details about the structure of a PDF, $\pdf(\bz)$, given a large number of samples $\{\bz_i\}$; see, e.g., Refs.\ \ocite{Shirts13,Parth,Khand,Shell12,Peter09,Faller12,Muller-Plathe2002}. In such cases, a parametric approach based on MLE  may not be suitable because a functional model $\pdf(\bz|\phi)$ may not be known {\it a priori}. Analysis based on minimal assumptions of distributional form goes by the name of {\it nonparametric statistics}. Histogram estimation of $\pdf(z)$ is a well-known technique routinely used from this class of methods. Unfortunately, due to the necessity of choosing bin number and sizes, histograms suffer from subjectivity and difficult-to-quantify uncertainties. Kernel density estimation is another approach although, once again, there is the vexing question of ``bandwidth selection'' \cite{izenman:1991}. In lieu of these, here we consider an approach based on spectral reconstructions; we call this method spectral Monte Carlo (SMC). 

The key idea behind SMC is to represent an arbitrary, one-dimensional, probability density as a spectral expansion of the form
\begin{align}
\pdf(\bz) \approx \pdf_M(z) = \sum_{m=1}^M a_m \phi_m(\bz) \label{eq:spectraldecomp}
\end{align}
Here $\phi_m(\bz)$ are orthonormal basis functions, $a_m$ are mode coefficients, and $M$ is a mode cutoff.  Analogous to orthogonal unit vectors in a high-dimensional space, the basis functions satisfy an integral orthogonality condition
\begin{align}
\int_D {\rm d}\bz\,\, \phi_m(\bz) \phi_n(\bz)=\delta_{m,n} \label{eq:orthogonality}
\end{align}
In the context of Eq.~\eqref{eq:spectraldecomp}, the $\phi_m(z)$ generalize the notion of a  histogram bin insofar as the latter is equivalent to an {\it indicator function}, that is a function whose value is $1$ when $\bz_i$ falls within the bin and zero otherwise.  The main idea in using Eq.~\eqref{eq:spectraldecomp} is to pick a basis set matched to the problem in the sense that one expects the first few functions to have a shape that well approximates the PDF under consideration.  For our applications in which the PDFs look more or less Gaussian, the Hermite functions are a suitable choice. These functions are defined via
\begin{align}
\phi_m(z) = H_m(z)\exp\left(-\tfrac12 z^2\right)
\end{align}
where the $H_m(z)$ are the normalized Hermite orthogonal polynomials, see Chap. 18 in Ref.\ \ocite{NIST:DLMF}. We refer to Ref.\ \ocite{schwartz1967} for perhaps the earliest use of these functions for this task, and Ref.\ \ocite{PatroneRosch17} for a more recent discussion on basis sets.  Moreover, subject to smoothness conditions on $\pdf(\bz)$, there exist many basis sets for which Eq.~\eqref{eq:spectraldecomp} converges to the true PDF when $M\to \infty$. One need not worry too much about the exact correspondence between the $\phi_m(\bz)$ and $\pdf(\bz)$ \cite{PatroneRosch17}.

Given Eq.~\eqref{eq:spectraldecomp} and a choice of $\phi_m(\bz)$, one can formally compute the mode coefficients by invoking the orthogonality relationship \eqref{eq:orthogonality}.  Specifically, multiplying Eq.~\eqref{eq:spectraldecomp} by $\phi_n$ and applying Eq.~\eqref{eq:orthogonality} yields the first of the equations below
\begin{align}
a_m = \int_D {\rm d}\bz\,\, \pdf(\bz) \phi_m(\bz) 
    \approx \amBar := \frac{1}{N}\sum_{i=1}^{N} \phi_m(\bz_i) \label{eq:weights}
\end{align}
As the $\pdf(\bz)$ is unknown, the $a_m$ cannot be computed.  However, the set $\{\bz_i\}$ amounts to $|\{\bz_i\}| = N$ samples drawn from $\pdf(\bz)$.  Thus, as with the statistical mechanical expectations discussed previously in Eq.\ \eqref{eq:MCapprox}, we may approximate the $a_m$ by the Monte Carlo estimate on the right.

\begin{figure*}
\includegraphics[width=12cm]{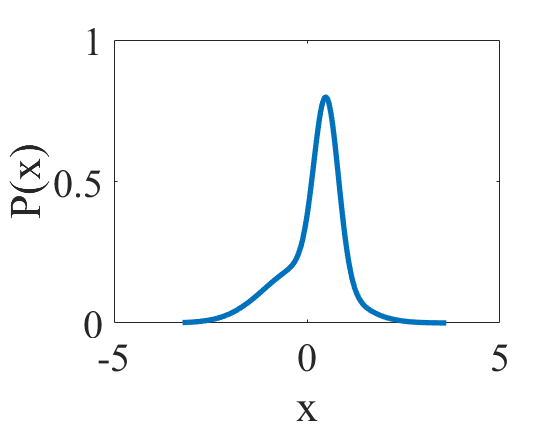}\caption{A ``duck-bill'' distribution corresponding to Eq.~\eqref{eq:samplepdf} with $a=b=1/2$, $\mu_1=0.5$, $\mu_2=-0.1$, $\sigma_1^2=0.1$, and $\sigma_2^2=1$.}\label{fig:duckbill}
\end{figure*}

\begin{figure*}
\includegraphics[width=16cm]{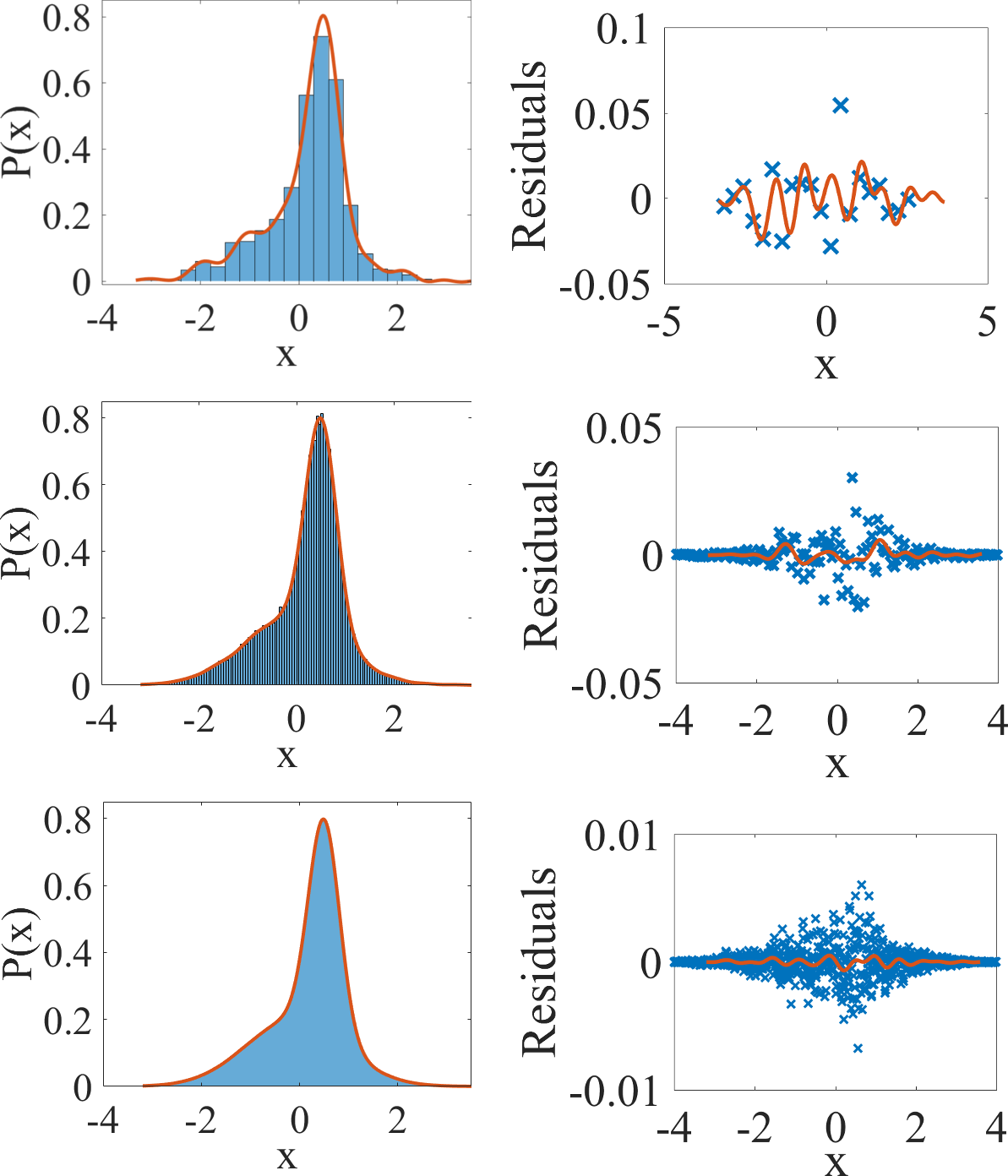}\caption{Comparison of SMC (orange curve) and a histogram (blue bins) for $N=1000$ (top), $N=10^5$ (middle), and $N=10^7$ (bottom row) realizations of a random variable drawn from Eq.~\eqref{eq:samplepdf} with the parameters used in Fig.~\ref{fig:duckbill}.  Residuals, i.e. the difference between a given reconstruction and the true density $\pdf(\bz)$ are shown on the right.  Histogram bin-widths are default values chosen by the Matlab function ``histogram.'' The analog to bin-width for SMC is truncation parameter. SMC yields an estimate of the pdf as a smooth function of the variable $\bz$.}\label{fig:comparison}
\end{figure*}

To illustrate the usefulness of SMC independent of any MD application, we invoke Eq.~\eqref{eq:spectraldecomp} to model a probability density of the form
\begin{align}
\pdf(\bz) = \frac{a}{\sqrt{2\pi\sigma_1^2 }}e^{-\frac{(\bz-\mu_1)^2}{2\sigma_1^2}} + \frac{b}{\sqrt{2\pi\sigma_2^2 }}e^{-\frac{(\bz-\mu_2)^2}{2\sigma_2^2}}\label{eq:samplepdf}
\end{align} 
where $a+b=1$, and $\mu_j$ and $\sigma_j$ are free parameters for $j=1,2$.  We pick default values of $a=b=1/2$, $\mu_1=0.5$, $\mu_2=-0.1$, $\sigma_1^2=0.1$, and $\sigma_2^2=1$, which leads to a ``duck-bill'' probability density illustrated in Fig.~\ref{fig:duckbill}.  Using a random-number generator, the script realizes $N$ independent samples of $\bz_i$ drawn from $\pdf(\bz)$ defined above and generates PDFs both as a histogram and in terms of Eq.~\eqref{eq:spectraldecomp}.  For illustrative purposes, Fig.~\ref{fig:comparison} compares these two methods for $N=1000$, $N=10^5$, and $N=10^7$.  Several observations are notable: (i) SMC produces smooth reconstructions on the domain of $x$, whereas the histogram is discontinuous; (ii) SMC yields consistently smaller residuals than a histogram; and (iii) as a consequence, requires less data to generate acceptable reconstructions.

\begin{figure}
\includegraphics[width=10cm]{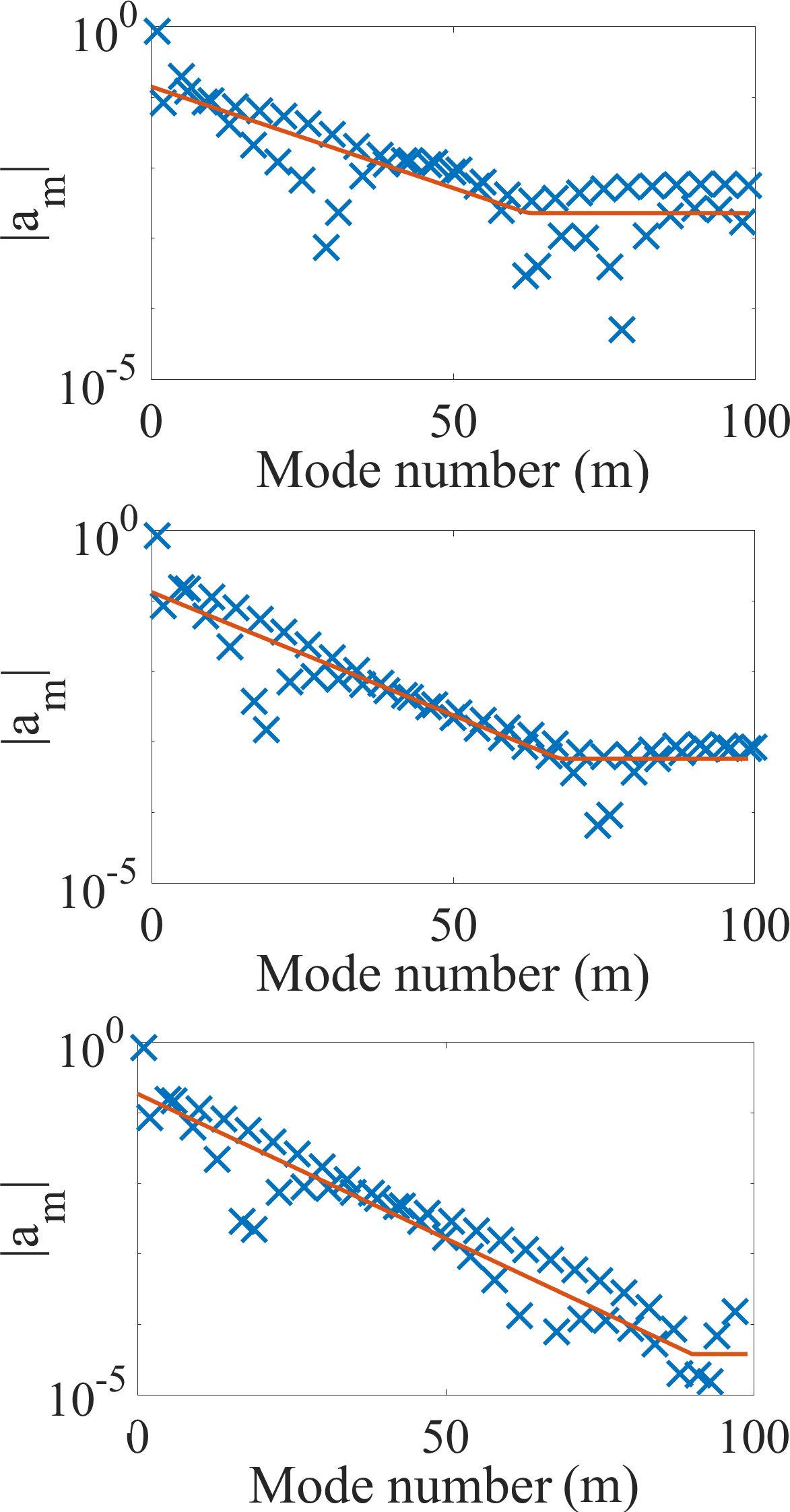}\caption{Absolute value of mode-weights $|a_m|$ (blue $\times$) as a function of mode number $m$ for $N=1000$ (top), $N=10^5$ (middle), and $N=10^7$ (bottom).  The orange lines are bilinear fits to $\log|\amBar|$ and serve as guides for the eye.  The point at which the slope of the bilinear fit changes is a reasonable estimate for the mode cutoff $M$.  Note that the noise-floor drops by roughly an order of magnitude for each factor of 100 increase in the number of samples. }\label{fig:modeweights}
\end{figure}

We end the discussion with three issues related to the implementation of SMC.  First, we note that in order for the spectral expansion Eq.~\eqref{eq:spectraldecomp} to be meaningful, the $a_m$ must decay as a function of $m$, assuming we can compute Eq.~\eqref{eq:weights} exactly.  However, the $\amBar$ as described in Eq.~\eqref{eq:weights} are random quantities that depend on the samples $\bz_i$.  As such, their values come with some uncertainty, which we can estimate via the standard error (cf.\ Eq.~\eqref{eq:standerr})
\begin{align}
\sigma_m^2 = \frac{1}{ N ( N -1)} \sum_{i} \big(\amBar - \phi_m(\bz_i)\big)^2 \label{eq:SMCerror}
\end{align}
In light of the decay of $a_m$, there should be an index $M$ for which $\amBar < \sigma_m$ for all $m > M$.  Because the signal is dominated by noise for such $\amBar$, they will be indistinguishable from the noise on a plot of $\log|\amBar|$ versus $m$.  We can thus identify a reasonable mode cutoff as the beginning of the noise floor, which is illustrated in Fig.~\ref{fig:modeweights}. 

Second, it is important to be aware that SMC reconstructions may be unusable and/or nonphysical in the tails of the distribution.  Because such tail events are rare, few realizations $\bz_i$ tend to sample the associated regions of $\pdf(\bz)$.  Thus, SMC reconstructions are heavily dominated by the behavior of the  $\phi_m(\bz)$ and may not be accurate far into the tails.  Perhaps worse, orthogonal functions are oscillatory and will therefore result in  negative (but small) probabilities when $\pdf(\bz)$ approaches zero.  This is seen, for example, in the top-left subplot of Fig.~\ref{fig:comparison}.

Third, SMC should be viewed as complementary to MLE as the two are best used in different situations.  The former tends to work well under data rich conditions, as can be seen from Eq.~\eqref{eq:SMCerror}.  That is, uncertainty in the mode weights scales as $1/N$, so that we expect the point-wise error in the PDF to be bounded by approximately $\mathcal O(M/N)$.  As Fig.~\ref{fig:comparison} and \ref{fig:modeweights} illustrates, this can be on the order of several percent even when $N=1000$.  Under such circumstances, MLE might be more appropriate if the modeler has {\it a priori} knowledge about the form of the PDF.   

\subsubsection*{Uncertainty propagation}

In many practical applications, the probability density $\pdf(\bz)$ may  not be of intrinsic interest if we wish to model uncertainties in some function $f(\bz)$.  In the context of MD, this occurs, for example, when a simulation computes data that must be analyzed to yield a final property prediction, and $f(\bz)$ corresponds to the data analysis routine itself.  In such cases, we therefore wish to {\it propagate} uncertainty from $\bz$ into $f(\bz)$.  We now consider this problem in detail.

Generally speaking, there are a variety of techniques for uncertainty propagation, depending on details of both the raw data and the function $f(\bz)$; see for example, Ref.\ \ocite{smith2013}.  Here we make a few assumptions that apply to many problems in MD.  For one, we assume knowledge of a reasonable estimate for $\pdf(\bz)$, which can be computed according to the inference methods discussed in the previous section.  Moreover, we assume that $f$ is relatively inexpensive to compute numerically, although it might be intractable analytically.  This situation arises, for example, when $f$ is a nonlinear function of some dataset $\{z(T_1),\ldots,z(T_N)\}$, so that, e.g.\ Gaussian random variables are transformed into something more complicated.  

Given these assumptions, we invoke a straightforward technique based on generating {\it synthetic datasets}; see also Ref.\ \ocite{hesterberg:2011}. The key idea of our analysis is to use $\pdf(\bz)$ to generate large numbers (e.g.\ $\mathcal O(10^5)$) of synthetic realizations $\tilde \bz_j$ via random number generators, where $j$ indexes the synthetic sets.  Given that $f(\bz)$ is inexpensive to compute, we directly propagate the collection of synthetic datasets through this function to generate a new set $\{ f_j\}$.  Then, we can either estimate the moments (e.g.\ mean and variance) of $f_j$ directly, or perform more sophisticated inference analyses using any of the methods described previously.

\begin{figure}
\includegraphics[width=16cm]{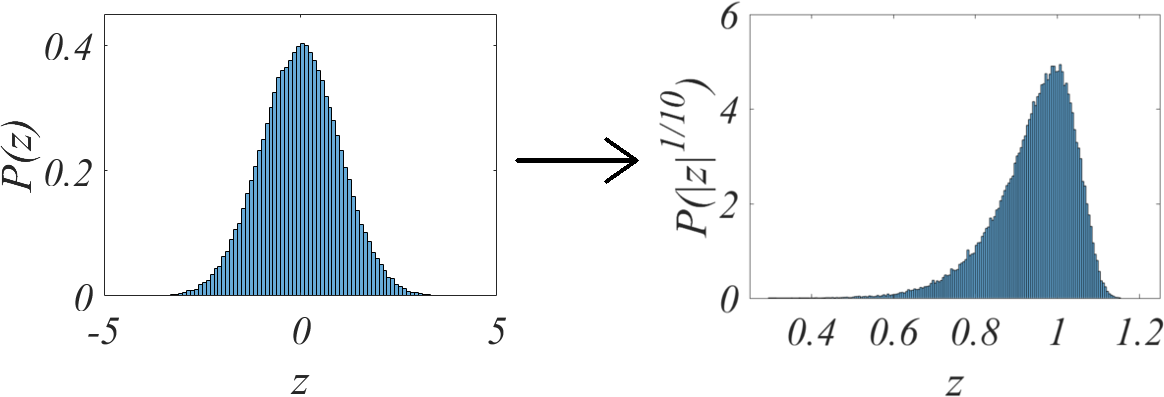}\caption{Histogram of $10^5$ points $z_i$ drawn from a standard normal random variable (left) propagated through the function $|z|^{1/10}$ (right).  Evaluation of this function is fast, so that we can approximate the properties of $P(|z|^{1/10})$ in terms of the samples shown on the right.}\label{fig:UCprop}
\end{figure}

To illustrate this idea, we consider an example wherein $\pdf(z)$ is a Gaussian PDF with zero mean and unit variance and $f(z)=|z|^{1/10}$.  Clearly $\pdf(f(z))$ is difficult to compute analytically, but we anticipate that it should have a peak near $f(z)=1$.  Figure \ref{fig:UCprop} illustrates the outcome of propagating $10^5$ realizations $z_i$ through this function, confirming our suspicion.  Using these realizations $f_i=f(z_i)$, we can estimate the statistical properties of $P(f(z))$.

\section*{UQ of MD}
\label{sec:UQofMD}
Having separately reviewed necessary ideas underlying MD and UQ, we are now in a position to combine these topics.  We frame this discussion in the context of three tutorials, which, roughly speaking, test and quantify uncertainties associated with a natural hierarchy of assumptions in MD: (i) discretization of Hamilton's equations; (ii) specification of ensemble degrees of freedom; and (iii) small system-sizes and times.

\subsection*{Tutorial: Trajectory Analysis}
\label{subsec:trajecsec}

As solutions $\bx(t)$ to Newton's equations are critical for dynamical sampling via Eq.~\eqref{eq:MCapprox}, the following questions arise naturally: (i) to what extent do discrete time-step algorithms reproduce these solutions; and (ii) how much uncertainty do they induce in simulated predictions?  Importantly, these questions are distinct insofar as the first amounts to asking whether a discrete solution $\bx_j$ approximates $\bx(jh)$ for all $j$, whereas the latter only asks whether the sets $\{\bx_j\}$ and $\{\bx(jh)\}$ are representative collections from the same $\pdf(\bx)$. As we show, the answer to the first question is unequivocally negative, $\bx_j \ne \bx(jh)$. By contrast, answers to the potentially more important second question appear more optimistic. Rather than discuss these questions in the abstract, we choose to examine them in the context of a harmonic oscillator. The elementary nature of this system is such that all aspects may be solved analytically. The hope is that this complete picture will provide an intuitive understanding of the theory for the general, intractable case.

\subsubsection*{Background}
The Hamiltonian for a single particle in one dimension (1D) is 
\begin{align}
 H_0( p, q)=\frac{ p^2}{2m} + \frac{m  \omega^2  q^2}{2} \label{eq:harmham}
\end{align}
where $m$ is the mass of the particle and $ \omega$ is the frequency.  For simplicity, we consider the NVE ensemble in which case $\pdf(\bx)$ is the uniform density on the $(p,q)$ ellipse defined by $H_0(p,q)=E$.  The equations of motion are
\begin{align}
\begin{pmatrix} \dot{q} \\ \dot{p} \end{pmatrix} 
  = \begin{pmatrix} 0 & 1/m \\ -m\omega^2 & 0\end{pmatrix}
    \begin{pmatrix} q \\ p \end{pmatrix} 
\end{align}
The linearity of these expressions has the fortunate consequence that all of the abstract machinery above---discrete timesteppers and symplectic analyses---may be discussed within the simpler context of linear algebra.  As a first exercise, we leave it to the reader to show that the continuous trajectory corresponding to Eq.~\eqref{eq:harmham} is given by the expression
\begin{align}
\binom{q(t)}{p(t)}= \matr
  {\cos(\omega t)}
  {\sin(\omega t)/(m\omega)}
  {-m \omega \sin(\omega t)}
  {\cos(\omega t)}
\binom{q_0}{p_0} = \bM_{c}(t)\binom{q_0}{p_0} \label{eq:harmsol}
\end{align}
where $\bM_{c}(t)$ is the continuous transition matrix that evolves initial data forward in time. Using trigonometric identities, it is easy to show the solution conserves energy as required, $E(t)=E(0)=p_0^2/2 + m\omega^2 q_0^2/2$. In the notation of the section on Hamilton's equations, fixing $t=h$ results in the exact flow map, $\Fh=\bM_c(h)$.

Analogous to Eq.~\eqref{eq:harmsol}, the discrete trajectories for the two integration routines previously discussed may also be computed explicitly as matrices. Concerning Velocity-Verlet, equations \eqref{eq:VV1}--\eqref{eq:VV3} result in
\begin{subequations}
\begin{align}
p_{n+\frac12} & = p_n - \frac{h}{2}(m\omega^2 q_n)\\
q_{n+1} &= q_{n}+h (m^{-1}p_{n+\frac12})\\
p_{n+1} &= p_{n+\frac12} -\frac{h}{2}(m\omega^2 q_{n+1})
\end{align}
\end{subequations}
Eliminating the intermediate term $p_{n+\frac12}$ and collecting coefficients of $(q_n,p_n)$ results in the VV transition matrix
\begin{equation}
\bMVV(h)
   = \matr
      {1-\frac{h^2\omega^2}{2}}
      {h/m}
      {-hm\omega^2\left(1-\frac{h^2\omega^2}{4}\right)}
      {1-\frac{h^2\omega^2}{2}}\label{eq:vvtransition}
\end{equation}
Similarly, the RK4 transition matrix is found to be 
\begin{align}
\label{eqn:transitionRK4}
\bMRK(h)
  & = \matr
      {1-\frac{h^2\omega^2}{2}+\frac{h^4\omega^4}{24}}
      {(h/m)\left(1-\frac{h^2\omega^2}{6}\right)}      
      {-hm\omega^2\left(1-\frac{h^2\omega^2}{6}\right)}
      {1-\frac{h^2\omega^2}{2}+\frac{h^4\omega^4}{24}}
\end{align}
Given initial data $\bx_0$, sets $\{\bx_1,\bx_2,...,\bx_n\}$ can then be computed in terms of the matrix powers $\bx_j = \bM^j \bx_0$, where $\bM$ stands for either $\bMVV$ or $\bMRK$. Again referring to the notation of the section on Hamilton's equations, either of these matrices represent a discrete time step map $\Dh$. Recall that, in general, the maps $\Fh$ and $\Dh$ are non-linear horrors whose existence is nevertheless dictated by theory. Here we prefer to keep the more friendly linear algebraic notation.

As a first pass at trajectory comparison we apply the local error analysis of Eq.~\eqref{eq:localerror}.  Expanding $\bM_c(h)$ as a Taylor series in $h\omega$ results in
\begin{subequations}
\begin{align}
\bMVV(h) - \bM_c(h)&= \frac{(h\omega)^3}{6}\matr{0}{(m\omega)^{-1}}{-\frac{m\omega}{2}}{0} + O((h\omega)^4)   \label{eq:vvdiff} \\
\bMRK(h) - \bM_c(h)&= \frac{(h\omega)^5}{120}\matr{0}{-(m\omega)^{-1}}{m\omega}{0} + O((h\omega)^6) \label{eq:rk4diff}
\end{align}
\end{subequations}
Taking the norm results in the bounds
\begin{subequations}
\begin{align}
\|(\bMVV(h) - \bM_c(h))\bx_0 \| <  \mathcal O\big((h\omega)^3\big) \\
\|(\bMRK(h) - \bM_c(h))\bx_0 \| <  \mathcal O\big((h\omega)^5\big)  
\end{align}
\end{subequations}
independent of $\bx_0$ as required by Eq.~\eqref{eq:localerror}. Furthermore, fixing an integration time $T$ and assuming that errors are cumulative over $\mathcal O(T/h)$ timesteps yields the estimates $\mathcal O(h^2 \omega^2)$ and $\mathcal O(h^4\omega^4)$ for upper bounds on the errors of the VV and RK4 algorithms, respectively.  That is, the former is a 2nd-order accurate method whereas the latter is 4th-order. From this perspective, Runge-Kutta would appear to be the clear algorithm of choice as it is higher order and, therefore, more accurate. It turns out that this assessment is too cursory in MD contexts and, to the extent that we know, MD software universally integrates the Hamiltonian system using Velocity-Verlet.

To better understand this last point, recall that we are considering here a system in the NVE ensemble.  Thus, one should assess the extent to which the $\bx_n$ sample the corresponding distribution $\mathcal P(\bx)$. In part, this amounts to verifying energy conservation.  While neither of the time-step algorithms conserve the original Hamiltonian, there is, nevertheless, a sharp distinction between $\bMVV$ and $\bMRK$ in this regard. Defining the perturbed Hamiltonian
\begin{equation}
  \tilde{H}(q,p;h) :=  \frac{p^2}{2m} + \frac{m\omega^2 q^2}{2} (1-h^2\omega^2/4)
\label{eq:HVV}
\end{equation}
it is straightforward to show that $\tilde{H}(q_{n+1},p_{n+1};h)=\tilde{H}(q_n,p_n;h)$ when using $\bMVV(h)$ to advance the system.  Notably, Eq.~\eqref{eq:HVV} is perturbation of $\mathcal O(h^2)$ relative to its original counterpart.

By contrast, one may show that $\bMRK(h)$ does not conserve any quadratic energy.  To see this, note that a general quadratic form can be expressed as
\begin{align}
H(p,q)=\bx^{\rm T} \bHam \bx \label{eq:matrixham}
\end{align}
where 
\begin{align}
\bHam=\begin{pmatrix}
A & B \\ B & C
\end{pmatrix}
\end{align}
is a $2\times 2$ symmetric matrix. In this representation, energy conservation amounts to the matrix equality
\begin{align}
\bMRK^{\rm T} \bHam \bMRK = \bHam \label{eq:renormham}
\end{align}
We leave it to the reader to show that this leads to a system of equations of the form
\begin{align}
\begin{bmatrix}
1-\alpha^2 & -2\alpha\gamma & -\gamma^2 \\
-2\alpha \beta & 1-\alpha^2 - \beta \gamma & -2\alpha \gamma \\
-\beta^2 & - 2\alpha \beta & 1-\alpha^2
\end{bmatrix} \begin{bmatrix}
A\\B\\C
\end{bmatrix}=0 \label{eq:RK4consistency}
\end{align}
where $\alpha= 1-\frac{h^2\omega^2}{2}+\frac{h^4\omega^4}{24}$, $\beta=\frac{h}{m}\left(1-\frac{h^2\omega^2}{6}\right)$, and $\gamma=-hm\omega^2\left(1-\frac{h^2\omega^2}{6}\right)$.  Mathematically, a non-trivial solution to Eq.~\eqref{eq:RK4consistency} exists if and only if the determinant of the matrix is zero. An elementary (but tedious!) computation shows this not to be so for general $h$, $m$, and $\omega$.  Thus, if RK4 does conserve an energy, it cannot be quadratic.

In practice, the existence of a conserved energy alone is not sufficient to guarantee that a given algorithm adequately samples phase space. It is here that the  symplectic analysis of time step algorithms becomes useful. One first needs to determine if either of the algorithms are symplectic, i.e.~satisfy Eq.~\eqref{eq:symplectic}. As the flow maps of this problem are linear, the Jacobians of the maps are simply the constant transition matrices $\bMVV(h)$ and $\bMRK(h)$. We leave it as an exercise to show that $\bMVV(h)$ satisfies,
\begin{align}
\bM^T\bj\bM = \bj
\end{align}
whereas $\bMRK(h)$ does not. Considering the symplectic Velocity-Verlet algorithm, from equation~\eqref{eq:vvtransition} we see that for small $h$, $\bMVV(h)$ is approximately the identity. Thus the conditions of Theorem~\ref{thm:shadowH} are satisfied. It follows that there exists a modified ($h$-dependent) Hamiltonian system such that the continuous trajectory sampled at times $t_n=nh$ result in exactly the same phase space points computed from the Velocity-Verlet time steps applied to $H_0$.

This modified Hamiltonian is sometimes referred to as a ``shadow Hamiltonian'' and, in the general case, its existence is taken as an indicator that the Velocity-Verlet algorithm is more appropriate for the problems at hand. For the sake of completeness (and because we can in this simple case) the shadow Hamiltonian may be computed explicitly.  Observing that $\bMVV(h)$ conserves the modified energy shown in Eq.~\eqref{eq:HVV} suggests that the shadow system may again be a harmonic oscillator, albeit with different, $h$-dependent mass and frequency: $m'(h)$ and  $\omega'(h)$. Substituting these unknown functions into Eq.~\eqref{eq:harmsol}, setting the time step to $h$, and requiring exact agreement with the VV map results in the system
\begin{equation}
\matr
  {\cos(\omega' h)}
  {\sin(\omega' h)/(m'\omega')}
  {-m' \omega' \sin(\omega' h)}
  {\cos(\omega' h)}
=\matr
  {1-\frac{h^2\omega^2}{2}}
  {h/m}
  {-hm\omega^2\left(1-\frac{h^2\omega^2}{4}\right)}
  {1-\frac{h^2\omega^2}{2}}
\label{eq:VVansatz}
\end{equation}
At first sight the situation seems hopeless as we have two unknowns to satisfy three equations. However, with judicious use of trig identities one finds that equations \eqref{eq:VVansatz} are consistent with the solution given by
\begin{subequations}
\begin{align}
\omega'(h) &= h^{-1}{\rm acos}\left(1-\frac{h^2\omega^2}{2} \right) \label{eq:renormomega} \\
m'(h)&=\frac{\omega m \sqrt{1-h^2\omega^2/4}}{\omega'(h)} \label{eq:renormmass}
\end{align}
\end{subequations}
In summary, for a given $h$, we define the shadow Hamiltonian by \eqref{eq:harmham} with $\omega'(h)$ and $m'(h)$ given by \eqref{eq:renormomega} and \eqref{eq:renormmass}. For any initial condition, the collection of phase space points resulting from integrating the original Hamiltonian system using the Velocity-Verlet algorithm corresponds exactly with the phase space points of the shadow system sampled at $t_n=nh$. Concerning the relationship between the renormalized Hamiltonian and the original problem, one finds that
\begin{align}
H'(p,q;h)=\frac{{\rm acos}\left(1-\frac{h^2\omega^2}{2} \right)}{h\omega\sqrt{1-h^2\omega^2/4}} 
   \left(\frac{p^2}{2m} + \frac{m\omega^2 q^2}{2} (1-h^2\omega^2/4) \right) = H_0(p,q) + \mathcal O((h\omega)^2)
\end{align}
Thus all aspects of the theory are confirmed.

We conclude this section with a few comments. In the general case the shadow Hamiltonian cannot be found. However, assuming that this object is a perturbation of the original system likewise suggests that the NVE manifolds are analogous. Admittedly this last point is a leap of faith given the high dimensionality and complexity of the systems under consideration. Nevertheless, practical experience confirms that the energy conservation properties of the 2nd-order Velocity-Verlet algorithm are significantly better than the 4th-order Runge-Kutta. This explains the widespread use of the VV algorithm despite its lower order.

Finally, we emphasize that the results discussed herein apply primarily to NVE or constant energy ensembles.  Thermostats such as Nos{\' e}-Hoover were developed in an effort to preserve the Hamiltonian framework while generating phase-space points from a canonical (NVT) ensemble. In this case, conservation of energy is not expected. However, the distribution of energy does have a structure predicted by statistical mechanics. A simple test to confirm correspondence between discrete trajectories and this theoretical distribution are discussed in the next section.

\subsection*{Tutorial: Ensemble Verification}
\label{subsec:shirtswork}

In the previous section, we considered how the symplectic structure of certain discrete integrators ensures that the microcanonical (or NVE) probability density $\pdf(\bx)$ is sampled correctly for a harmonic oscillator.  Here, we extend the discussion to ensemble verification of systems that are specified in terms of a constant temperature; generalizations to constant pressure simulations are straightforward.  The bulk of this section uses methods originally proposed by Shirts \cite{Shirts13}, although we modify his analysis to incorporate tools that have recently been developed.

\subsubsection*{Background and main ideas}

In statistical mechanics, it is well known that the Boltzmann factor associated with \changepp{the}  probability of a microstate $\bx$ can be rewritten in terms the system energy $E$ alone.  In particular, one finds
\begin{align}
\pdf(E|T) = \int\!\!\!{\d}\bx \,\,\delta(E - \mathcal H(\bx))\pdf(\bx|T) = \frac{\Omega(E)}{\mathcal Z(T)} \exp\left(\frac{-E}{k_bT} \right) \label{eq:ebolt}
\end{align}
where $\Omega(E)$ is the density of states  and $\mathcal Z$ is the partition function.  Importantly, Eq.~\eqref{eq:ebolt} implies that for the canonical ensemble, we can drop the $\bx$ dependence when referring to states with the same energy.  Given the functional form for $\Omega(E)$, it would then be straightforward to verify Eq.~\eqref{eq:ebolt} holds for a given simulation, since it is easy to compute the energy for a simulated microstate.  Unfortunately, $\Omega(E)$ is formally defined via the expression
\begin{align}
\Omega(E)\propto \int_{H(\bx)=E} \!\!\!\!\!\!\!\d\bx\,\,\,\, \frac{1}{|\nabla_{\bx}\mathcal H(\bx)|} \label{eq:DOS}
\end{align}
where integration occurs over a constant energy surface and the omitted constant of proportionality amounts to a ``unit'' of phase space \cite{Pathria}.  Given that (i) the term $|\nabla_{\bx}\mathcal H(\bx)|$ may not be accessible from all simulation platforms and (ii) the constant energy surface $\{\bx:H(\bx)=E\}$ is typically intractable to compute, there is little hope of using Eq.~\eqref{eq:DOS} in a practical setting.

Close examination of Eq.~\eqref{eq:DOS}, however, reveals a fortunate and useful fact: while $\Omega(E)$ has a complicated and intractable dependence on system parameters, it does not depend on temperature.  This suggests eliminating $\Omega(E)$ by considering the ratio
\begin{align}
\frac{\pdf(E|T_1)}{\pdf(E|T_2)} = \frac{\mathcal Z_2}{\mathcal Z_1}\exp\left( -E \left(\beta_1 - \beta_2 \right) \right) \label{eq:DOSratio}
\end{align}
where $\beta_i = 1/(k_b T_i)$ are the inverse temperatures.  Taking a logarithm yields
\begin{align}
R(E)=\log \left(\frac{\pdf(E|T_1)}{\pdf(E|T_2)} \right) = (\beta_1 A_1 - \beta_2 A_2) - (\beta_1 - \beta_2)E \label{eq:shirtslinear}
\end{align}
where $A_i$ are the Helmholtz free energies \cite{Pathria}.  The point is that Eq.~\eqref{eq:shirtslinear} is a linear function of $E$ with a slope that depends on the difference of reciprocal temperatures.  As such, it should be possible verify this slope by running simulations at two different temperatures, constructing $\pdf(E|T_1)$ and $\pdf(E|T_2)$, and computing a linear regression to the log of their ratio \cite{Shirts13}.  In the following section, we discuss a procedure for doing this in the context of provided tutorial scripts.

\subsubsection*{Example: application to water simulations}

To illustrate a verification test based on Eq.~\eqref{eq:shirtslinear}, we consider TIP4 water simulations using different thermostats \cite{Jorge}; results may be reproduced with provided files.  %
Given a user-specified temperature $T$ and thermostat,  we first run an NVT simulation for 100 ps (1 fs timestep) in order to equilibrate the system to its current temperature.  Next, we run a 2 ns simulation with a 1 fs timestep, outputting the energy every 100 timesteps.  These values are saved to a file that is subsequently analyzed by a Matlab script.  We denote these energies $\{E_i(T)\}$ for $1\le i \le 20000$.

Given two such files generated at different temperatures $T_1$ and $T_2$, the Matlab script next estimates probability densities $\tilde {\pdf}(E|T_1)$ and $\tilde {\pdf}(E|T_2)$ using the spectral Monte Carlo method discussed in the previous section.  As a preliminary step, we compute the sample mean $\bar E(T)$ and variance $\sigma_E(T)$ for each set $\{E_i(T)\}$ according to Eqs.~\eqref{eq:sampmean}-- \eqref{eq:sampvar} and rescale the energies via 
\begin{align}
\epsilon_i(T) = \frac{E_i(T) - \bar E(T)}{\sigma_E(T)} \label{eq:energyscale}
\end{align}
where $i$ indexes the realization of energy output by the simulation and $\bar E(T)$ is the sample mean, which is recomputed for each temperature\blfootnote{When estimating PDFs using SMC, this step often facilitates rapid convergence because the Hermite functions are centered around the origin.}.  Next, taking $\phi_m(\epsilon)$ to be Hermite functions, we approximate
\begin{align}
\pdf(\epsilon|T) \approx \tilde {\pdf}(\epsilon|T)= \sum_{m=1}^M \bar a_m \phi_m(\epsilon) \label{eq:PSMC}
\end{align}
where the $\bar a_m$ are computed according to Eq.~\eqref{eq:weights} using the $\{\epsilon_i(T)\}$.  Here the mode cutoff $M$ is chosen according to the method described in the SMC section.  Rescaling the resulting PDF in terms of $E$ (i.e.\ using Eq.~\eqref{eq:energyscale}) yields a smooth estimate of $P(E|T)$ that we can evaluate on a continuous domain.

\begin{figure}
\includegraphics[width=16cm]{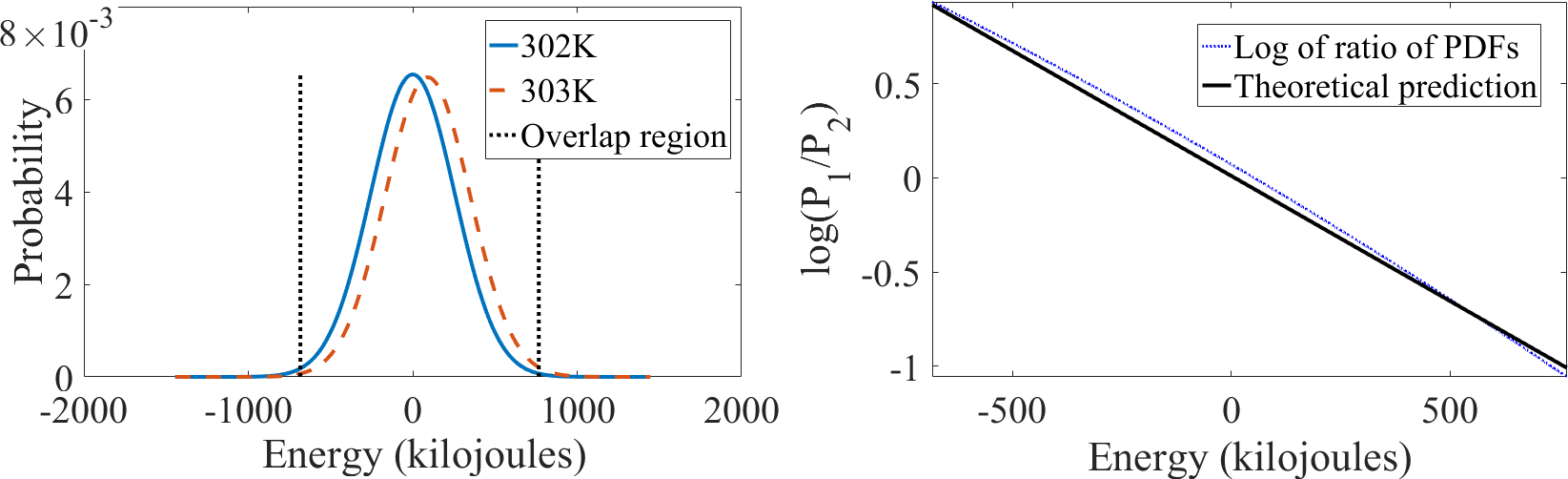}\caption{Left: Probability densities computed according to Eq.~\eqref{eq:PSMC} for two water simulations run at 302 K and 303 K and using the Nos{\' e}-Hoover thermostat.  Right:  The log-ratio of the PDFs compared against the theoretical prediction according to Eq.~\eqref{eq:shirtslinear}.  Close agreement between the slopes suggests that the thermostat is adequately sampling phase-space points from a canonical distribution.}\label{fig:logratios}
\end{figure}

\begin{figure}
\includegraphics[width=10cm]{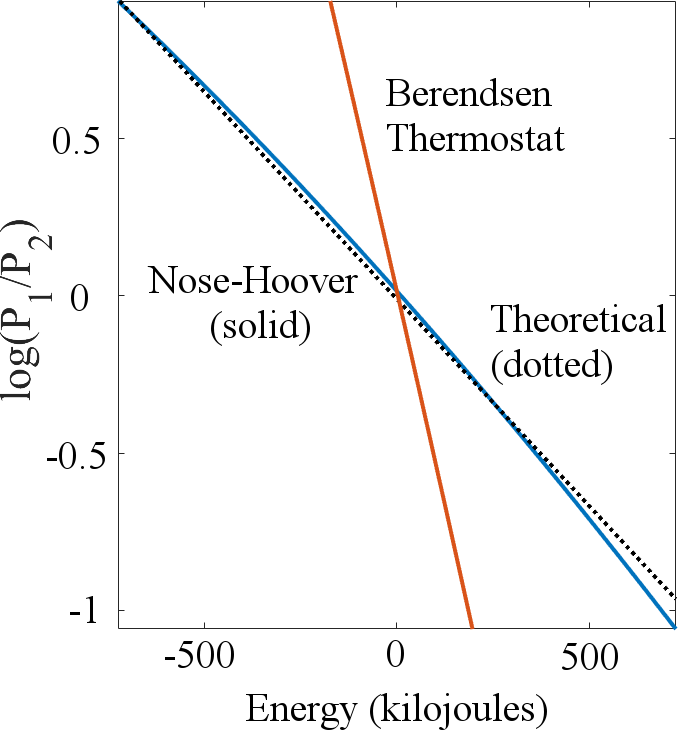}\caption{Comparison of Nos{\' e}-Hoover and Berendsen thermostats when $T_1=300$ K and $T_2=301$ K.  Note that the slope of the latter differs significantly from the theoretical prediction, indicating that this thermostat is not consistent with a canonical ensemble. }\label{fig:thermocompare}
\end{figure}

Given reconstructions $\tilde {\pdf}(E|T_1)$ and $\tilde {\pdf}(E|T_2)$ at two different temperatures, we next compute an overlap domain $D$ on which to perform Shirts' test.  In particular, we assume that the spectral reconstructions are accurate up to 3 standard deviations from their respective means, which yields $D=[\bar E(T_1) - 3\sigma_E(T_1),\bar E(T_1)+3\sigma_E(T_1)]\cap[\bar E(T_2)-3\sigma_E(T_2),\bar E(T_2)+3\sigma_E(T_2)]$.  Given this, it is straightforward to evaluate the log-ratio in Eq.~\eqref{eq:shirtslinear} on a dense grid of energies $E\in D$, which we denote
\begin{align}
\tilde R(E) = \log\left[\frac{\tilde {\pdf}(E|T_1)}{\tilde {\pdf}(E|T_2)} \right] \label{eq:RE}
\end{align}
when written in terms of the approximate PDFs.  Figure \ref{fig:logratios} shows an example of this analysis applied to two water simulations run at 302 K and 303 K using the Nos{\' e}-Hoover thermostat.  By eye, the analysis indicates good agreement between the slope of the simulated log-ratio and its predicted counterpart $m_{\rm theory}=\beta_1-\beta_2$, suggesting that the thermostat is consistent with a canonical ensemble.  Figure \ref{fig:thermocompare} compares the Nos{\' e}-Hoover and Berendsen thermostats when $T_1=300$ K and $T_2 = 301$ K.  While the former is consistent with the theoretical prediction, the slope associated with the latter deviates noticeably.   These results are consistent with Ref.\ \ocite{Shirts13}, as well as known results on the Berendsen thermostat \cite{Icecube,Icecube2}, indicating that it does not sample the Boltzmann distribution.  See Ref.~\ocite{Shirts13} for a more detailed discussion of other thermostats.  

Despite the potential usefulness of this analysis for identifying nonphysical thermostats, we emphasize that in its current state, it is limited insofar as agreement with Eq.~\eqref{eq:shirtslinear} does not imply that we have actually sampled a canonical distribution.  This could happen, for example, if the system becomes trapped in a local energy minimum but otherwise samples energies according to the Boltzmann distribution.  Moreover, we are not guaranteed that other physical properties (besides energy)  are appropriately sampled.  More generally, the ensemble verification test does not provide us with a definitive way to quantify the extent of sampling or relate this to associated uncertainties in predicted quantities.  Thus, Eq.~\eqref{eq:shirtslinear} is to a certain extent a qualitative tool, and care should be exercised in drawing conclusions based on its outcomes.

\subsection*{Tutorial: UQ of data analysis for the glass-transition temperature}

In this last tutorial, we demonstrate a UQ workflow that provides a final estimate of a material property, the glass-transition temperature $T_g$, along with its associated uncertainties.  This task represents a culmination of the chapter insofar as it demonstrates the role UQ plays in building confidence in models and simulation.  The hope is that, an industrial modeler could use this workflow as the basis for a decision to invest experimental resources in a material based on computational predictions of desired properties.  

We note that while the goal of this example is essentially to compute error bars, additional verification steps specific to $T_g$ may be needed  to ensure that our final estimates are meaningful in practice; see Ref.~\ocite{Patrone16}.  While we cannot provide general recipes for such verification tasks (given the diversity of material properties and methods of computing them), we emphasize that modelers should always take the time to assess the extent to which raw data coincides with physical and theoretical expectations.

\subsubsection*{Background and underlying ideas}

In polymer physics, the glass transition temperature is, roughly speaking, the temperature at which a material becomes soft \cite{Donth}.  Within the community, there is some ambiguity as to exactly what $T_g$ represents and how to measure it; thus a variety of empirical and experimental definitions have been proposed reflecting different approaches to clarifying the notion.  What is generally agreed upon, is that physical properties such as elastic moduli and density undergo a rapid transition between different behaviors near $T_g$. Thus, many approaches use this observation as the basis for defining $T_g$ in terms of experimental data.

\begin{figure}
\includegraphics[width=12cm]{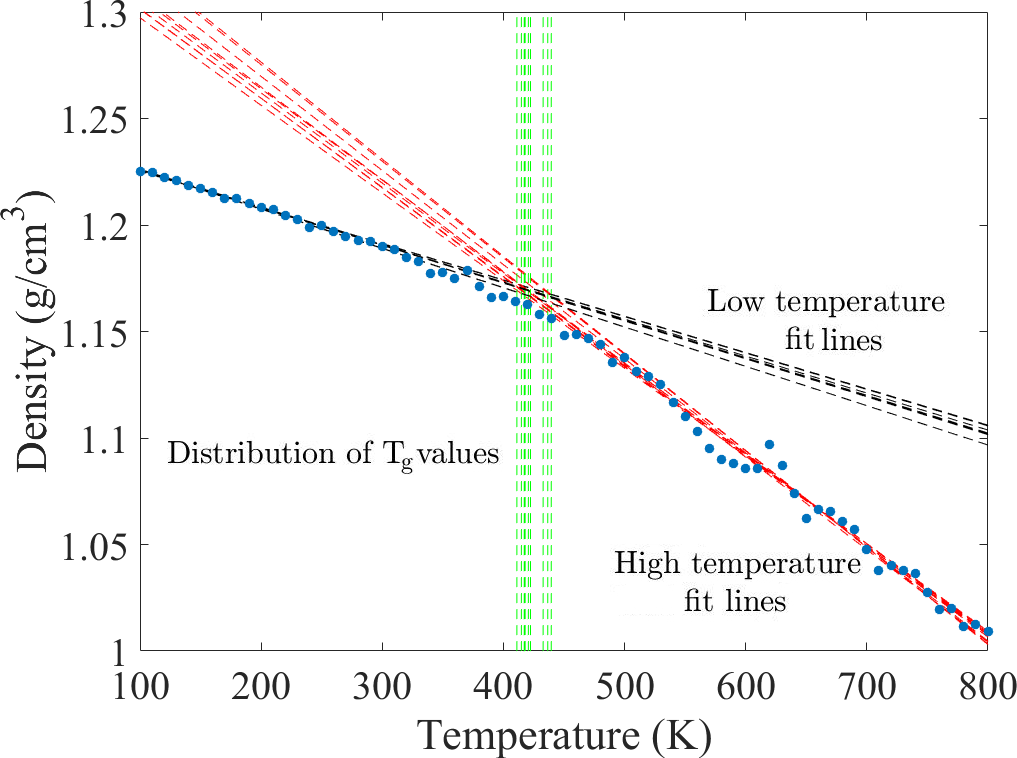}\caption{Typical simulated $\rho(T)$ data for a crosslinked polymer system; see the main text for discussion of the simulations.  Multiple fit lines illustrate that many plausible $T_g$ values could be extracted from this data.  In this example, the fit lines are 100 K and 800 K tangents to hyperbola fits of synthetic datasets generated according to the procedure described in the main text. }\label{fig:fitlines}
\end{figure}

Perhaps the simplest such definition equally accessible to both simulations and experiment arises from examination of density-vs-temperature curves, $\rho(T)$.  Experimentally, these data are generated by annealing a sample at a fixed rate (typically 10 K per minute) and measuring the density as a function of the temperature.  {\it In silico} it is straightforward to reproduce this procedure, albeit at significantly faster cooling rates (e.g., on the order of $10^9$ K/min), due to the timescale limitations of MD. In this scenario, the definition of glass transition is based on the observation that density varies linearly with temperature when the material is in its glassy and rubbery states, i.e.,~at asymptotically low and high temperatures respectively.  The glass transition is said to occur when the slope of $\rho(T)$ transitions rapidly between these asymptotic regimes.  Conventionally, $T_g$ is defined as the intersection of two best-fit lines extrapolated from the low and high temperature data, which should meet in the middle of the transition region; see Fig.~\ref{fig:fitlines} and Refs.\ \ocite{Alesimandexp,Koo14,Varshney08,Li10,Li11,Fan07,Yu09,Li13,Khare09,Khare12,Khare13}.  

\begin{figure}
\includegraphics[width=18cm]{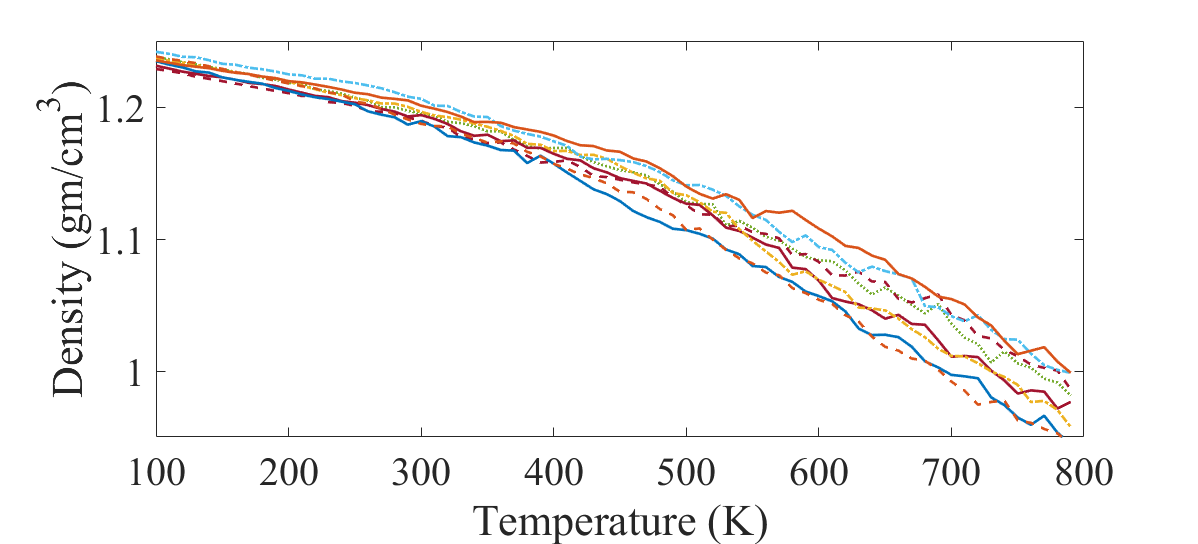}\caption{A collection of typical density-temperature curves extracted from separate simulations of the same system.  While all of the curves generally have the same shape, they nonetheless appear rotated and shifted relative to one another.  }\label{fig:noverlap1}
\end{figure}

In experiments, this transition is relatively sharp and occurs over a narrow range of temperatures, with correspondingly obvious asymptotic regimes. In such cases, there is little uncertainty associated with the data analysis {\it per se}.  In MD simulations, however, several problems arise.  For one, small systems exhibit fluctuations in data due to finite-size and thermal effects, leading to uncertainty in the slope of the corresponding linear behavior; see Fig.~\ref{fig:fitlines} and Fig.~3 in Ref.\ \ocite{Li11}.  Perhaps worse, the transition region between asymptotic regions can occur over hundreds of degrees, making it difficult even to identify asymptotic regimes in the first place.  These issues are further compounded by the fact different simulations may output significantly different datasets that neither overlap nor agree in their estimates of $T_g$; see Fig.~\ref{fig:noverlap1}.  

Conceptually, then, these observations motivate the following statistical model. We assume that simulated predictions for $T_{\rm g}$  can be written as
\begin{align}
\tgi = \TgBar + \tgwithin + \tgbetween \label{eq:mixedeffects}
\end{align}
where the subscript $i$ indexes the simulation, $\TgBar$ is an average value of $T_{\rm g}$ obtained from an idealized, infinitely large simulation, $\tgwithin$ is a {\it within-simulation} uncertainty associated with an analysis of a single dataset, and $\tgbetween$ is a {\it between-simulation} uncertainty.    Physically, the $\tgwithin$ accounts for thermal noise and finite-size effects associated with a single density-temperature curve (as illustrated in Fig.~\ref{fig:fitlines}), whereas $\tgbetween$ accounts for undersampling of the system morphology (described in the next section), which manifests as differences between datasets (as seen in Fig.~\ref{fig:noverlap}).
Our goal is to quantify both of these uncertainties and, ultimately, to estimate $\TgBar$, the simulated prediction of $T_{\rm g}$ for a bulk material, taking all uncertainty sources into account. Toward this end, the additive structure of the noise in Eq.~\eqref{eq:mixedeffects} complicates this process, since we cannot hope to infer $\tgwithin$ and $\tgbetween$ on the basis of the same information.  We therefore separate analyses of these quantities in the sections that follow.

Our UQ program for this tutorial is as follows.  First, we review necessary elements of the simulation scripts and consider in more detail the definition of $T_g$.  Second, we propose a method for estimating $\tgwithin$ given a \changeAD{single} dataset. \changeAD{Next, given a collection of $T_g$ estimates, we estimate the contribution of between uncertainty using a maximum liklihood framework. Finally, we compute a consensus mean estimate for $\TgBar$ and its associated uncertainty.} 

\subsubsection*{\changepp{Simulations and $T_g$ estimates}}
\label{subsubsec:scripts}

\begin{figure}
\includegraphics[width=16cm]{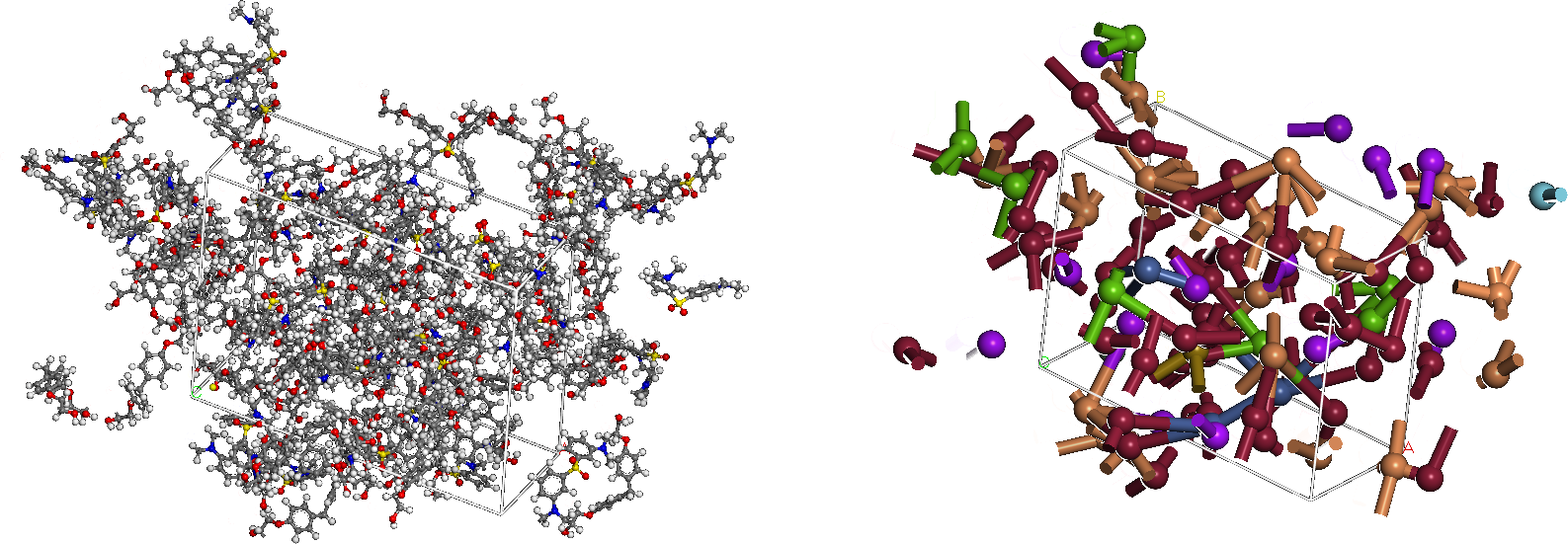}\caption{Example of a 33BF system.  The left figure shows an atomistic view of the unit cell, which contains a random network of crosslinked polymers.  The right figure shows a coarse-grained version of the same system in order to better resolved the network structure.  Beads correspond to epoxy and amine molecules; see Ref.~\ocite{Patrone16} for more details.} \label{fig:hairball}
\end{figure}

\changepp{To generate data for subsequent analysis, we run annealing simulations on an amine-cured epoxy (or thermoset polymer system),} which is commonly used in the aerospace industry.  For convenience, we refer to this system as 33BF.  Here we do not provide further details about the chemistry or how the systems were built; see Ref.~\ocite{Patrone16} (which uses the same abbreviation) \changepp{as well as the provided example scripts} for details.  However, we note that in bulk, this material is amorphous, so that the corresponding MD models are random networks of molecules; see Fig.\ \ref{fig:hairball}.  We interpret each such model as a different microscopic sample drawn from a bulk material.

The annealing script itself works as follows.  First, the specified system is heated to 800 K and allowed to equilibrate for 10,000 timesteps \changepp{(each of which is 1 fs)}, with densities output every 100 timesteps, i.e.\ corresponding to $N=100$ density samples.  \changeAD{(Should simulation parameters and timestep be reported in physical units?)} After completion of this simulation, the script enters a convergence-test loop, therein computing the first sample average $\tilde \rho_1$ and standard error $S_1$ via the formulas
\begin{align}
\tilde \rho_n &= \frac{1}{nN}\sum_{j=1}^{nN} \rho_j \label{eq:sampavg} \\
S_n^2 &= \frac{1}{nN(nN-1)} \sum_{j=1}^{nN} (\tilde \rho - \rho_j)^2 \label{eq:astanderr}
\end{align}
where $\rho_j$ is the $j$th density output by the script and the index $n$ refers to the iteration within the convergence loop, which starts at $n=1$.  Next, we compare $S_1$ with a user-defined equilibration convergence criterion, which is set by default to $c_e=0.001$ g/cm${^3}$.  If $S_1<c_e$, the simulation proceeds to the next step.  However, if $S_1>c_e$, the script iteratively performs 10,000 step simulations at the same temperature, incrementing $n$ by one each time.  After each simulation, $\tilde \rho_n$ and $S_n$ are computed, using the $nN$ available density samples, again checking the convergence criterion.  When $S_n< c_e$, the system is assumed equilibrated and the script exits the loop.

To compute an average density at \changeAD{this} temperature, the script first discards \changeAD{the density data obtained in the step above as it is considered transient}.  Then, it repeats the entire procedure described in the previous paragraph, but using a more stringent convergence criterion, which is set to $c_a=0.0001$ g/cm${}^3$ by default.  The final running average $\tilde \rho_n$ computed in this way is then recorded as the density output by the simulation at that temperature.  Moreover, the convergence parameter $c_a$ is an estimate of the uncertainty with which we know $\bar \rho(T)$ at any simulated temperature.  

To generate a density-temperature curve, the script decreases the temperature in 10 K increments, each time performing the above equilibration and averaging steps.  For the thermoset system considered here, we typically suggest annealing from 800 K to 100 K in order to sample the asymptotic temperature regimes required for $T_g$ estimation.  To allow the density to change with temperature, we use the Nos{\' e}-Hoover thermostat and barostat in LAMMPS \cite{Parrinello81,martyna:1994,tuckerman:2006,shinoda:2004}.  The collection of temperatures $(T_1,T_2,...,T_D)$ and densities $(\rho_1,\rho_2...,\rho_D)$ are denoted by $\T$ and $\brho$.

\begin{figure*}
\begin{center}
\includegraphics[width=16cm]{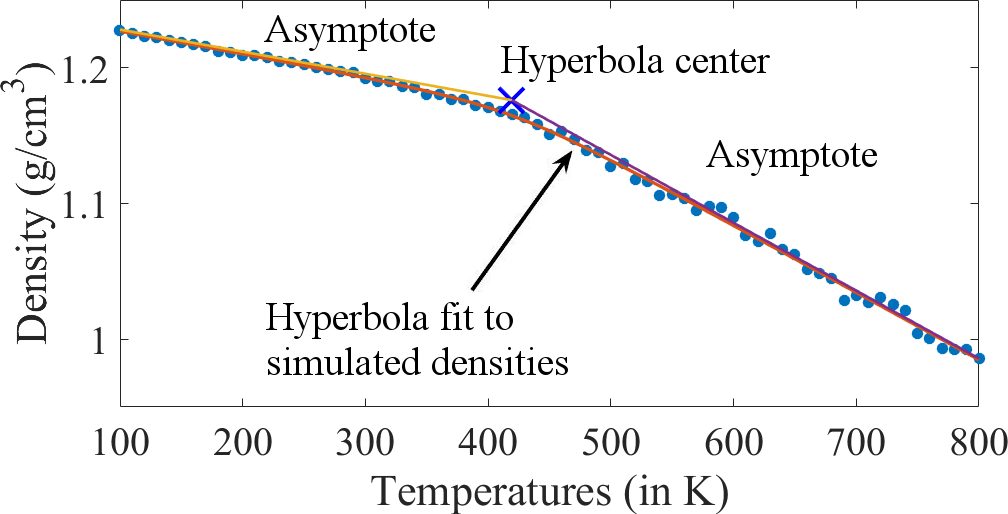}
\caption{A hyperbola (pink) fit to the 3800 atom 33BF data shown in Fig.~\ref{fig:fitlines}.  Here \Tg is given by the $T$ coordinate of the hyperbola center.  The hyperbola asymptotes are shown to verify that our definition is consistent with the bilinear fit method.  }\label{fig:hypfit} 
\end{center}
\end{figure*}

To estimate $T_{\rm g}$, we invoke a method based on a nonlinear regression to temperature-density data\blfootnote{The hyperbola approach benefits from improved stability relative to the bilinear fits, which in effect amount to extrapolation; see Ref.~\ocite{Patrone16}.}.  In more detail, we assume a hyperbola model these data using the parameterization
\begin{subequations}
\begin{align}
\rho_{\H}(T,\phi) &= \rho_0 - a (T - T_0) - b \H(T-T_0,c)  \\
\H(T,c) & = \frac{1}{2}T +  \sqrt{\frac{T^2}{4} + \e^{\mathit{c}}}
 \label{eq:hyperbola}
\end{align}
\end{subequations}
where $T_0,\rho_0,~a,~b,$ and $c$ are constant parameters to be determined \cite{Cologne}; see Fig.\ \ref{fig:hypfit}.  Geometrically, the role of these parameters can be understood as follows.  First, in the limit that $c\to -\infty$, the function $\H$ approaches a piecewise linear function that is identically zero for $T < T_0$ and has unit slope for $T > T_0$.  The parameter $c$ smooths out the discontinuity in the slope at $T=T_0$, and thus controls the transition between asymptotes.  The point $(T_0,\rho_0)$ is the hyperbola center.  The parameters $-a$ and $-(a+b)$ are the slopes of the asymptotic low and high-temperature regimes, respectively.  For our applications, we may assume that $\rho_0$ and $T_0$ are both positive, since these correspond to densities and temperatures.  Likewise, we expect that $a$ and $b$ are positive, as density is a concave function of temperature for polymer systems.  The parameter $c$ may take any value $-\infty < c < \infty$.   We denote the collection of these parameters \changeAD{by the} vector $\phi$, and the resulting density as a function of temperature by $\rhoH(T,\phi)$.

To determine $\phi$ given a dataset $(\T,\brho)$, we solve the non-linear least-squares problem
\begin{equation}
\phi_{\rm LS} = \argmin_{\phi} \sum_{j=1}^D \left(\frac{\rhoH(T_j,\phi)-\rho_j}{\sigma_j}\right)^2
\label{eq:rhoSolve}
\end{equation}
where $\sigma_j=T_j^p$ for some $p>0$.  We justify Eq.~\eqref{eq:rhoSolve} later, noting only that the residuals to such a fit appear to have a constant variance and be uncorrelated\blfootnote{We have found that the precise value of $p$ has little effect on the fits, provided $1/2 \lesssim p \lesssim 2$.  Moreover, Eq. \eqref{eq:rhoSolve} is equivalent to the maximum likelihood estimate of $\phi$ under the noise model \eqref{eq:rhopostulate} when $\brhowithin$ is uncorrelated Gaussian white noise with a variance that scales as $T^{2p}$.}.  While beyond our scope here, it is interesting to note that the parameter $c$ can be used to determine how close the hyperbola asymptotes are to the data, thereby providing a means to verify that the simulation exhibits asymptotic behavior; see Ref.~\ocite{Patrone16} for details.  

\subsubsection*{Within-uncertainty estimate for \Tg}
\label{subsec:within}

Recall that Eq.~\eqref{eq:mixedeffects} posits two sources of uncertainty, within and between.  The first of these is unique to each dataset and is taken to model thermal and finite-size effects in a given simulation.  Mathematically, we view $\tgwithin$ as uncertainty in the value of $T_{\rm g}$ arising from the fact that $\rho_{\H}$ is not a perfect representation of simulated data $(\brho,\T)$.  To assess this, we therefore invoke the noise propagation analysis described previously.

\begin{figure}
\begin{center}
\includegraphics[width=16cm]{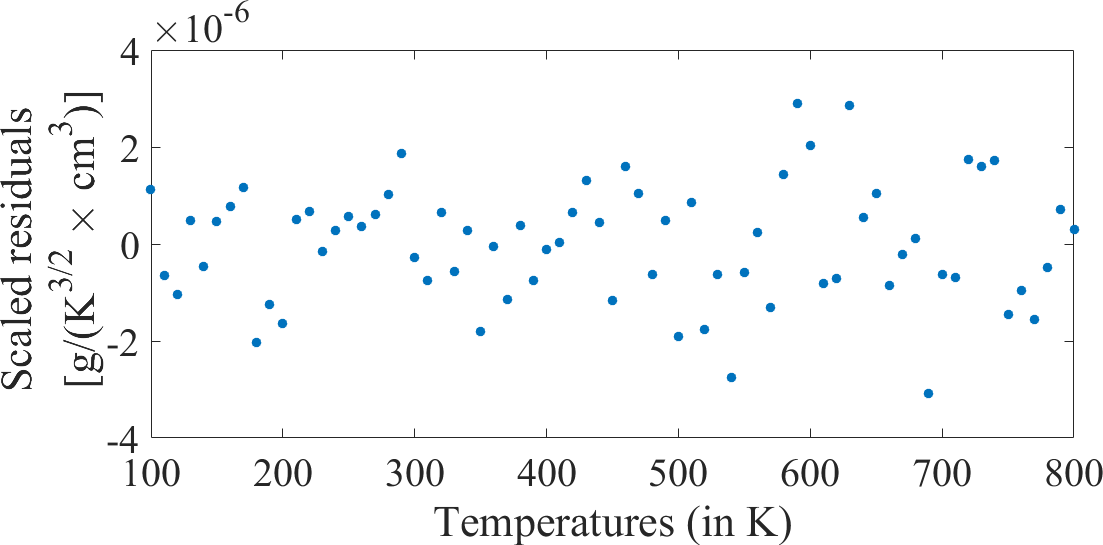}
\caption{Residuals $\rho(\T,\hat \phi) - \brho$ after rescaling by $\T^{5/4}$.  }\label{fig:scaledresids} 
\end{center}
\end{figure}

In more detail, let $(\T,\brho_i)$ denote a fixed dataset, with $\brho_i=(\rho_{i,1},...,\rho_{i,D})$ and $j$ ($1\le j \le D$) indexing data-points within the set.  In light of Eq.~\eqref{eq:mixedeffects}, the between-noise $\tgbetween$ appears as a fixed term that can be temporarily absorbed into $\TgBar$.  Thus we consider the reduced model
\begin{align}
\brho_i &= \rho_{\H}(\T,\phi_{\rm LS}) + \brhowithin \label{eq:rhopostulate} \\
\tgi & = \TgBari + \tgwithin
\end{align}
where $\rho_{\H}(\T,\phi)$ is the average behavior of the density data, $\brhowithin$ is a random vector that characterizes noise in a single dataset, and the notation  $\TgBari$ indicates that averages between datasets differ by virtue of the $\tgbetween$.  In this context, our task is therefore to estimate  $\varsigma_i^2 = \Var[\tgwithin]$ (which depends on the $i$th dataset) as a quantification of the within-simulation uncertainty in {\Tg} by propagating $\brhowithin$ through $\mathcal H$.

To achieve this, we postulate a multi-variate Gaussian model for $\brhowithin$ and infer the underlying parameters from the given dataset.  As reflected in Eq.~\eqref{eq:rhopostulate}, we interpret the residuals $\brho_i-\rhoH(\phi_{\rm LS},\T)$ as a sample of $\brhowithin$. Taking a fixed polymer system of modest size, we compute ten independent annealing runs and superimpose the residuals as a function of temperature.  A few observations can be made.  First, there is no discernible bias in the residuals.  This is significant as the hyperbolic fit, being a global model for $\rho(T)$ with relatively few parameters, imposes non-trivial structure on the computational data.  Departure from this structure would be observable as bias.  As we visually observe no bias we conclude that the hyperbola model is suitable and, furthermore, $\brhowithin$ may be assumed to have zero mean.  Second, as there are neither clear indications nor theoretical motivations for correlations between densities at different temperatures, we assume zero correlation, i.e.,~$\Cov(\brhowithin)={\rm diag}(\sigma_{i,j}^2)$; note that diagonalization occurs over $j$, which indexes temperature [cf.\ text under Eq.~\eqref{eq:rhopostulate}].  Finally, the residuals show increased noise at higher temperatures.  Intuitively this is reasonable from thermodynamic considerations, and furthermore motivates a power-law dependence
\begin{align}
\sigma_{i,j} = T_j^{p}\sigma_{*,i}  \label{eq:syntheticmodel}
\end{align}
Empirically, we find that setting $p=5/4$ yields scaled residuals that have approximately a uniform variance; see Fig.~\ref{fig:scaledresids}, which is representative of our computational datasets.  For $p$ fixed we estimate the overall scale of the variance by
\begin{align}
\sigma_{*,i}^2 := \frac{1}{D}\sum_{j=1}^{D} \frac{ [\rho( \phi_{\rm LS},T_{j}) - \rho_{i,j}]^2}{T_{j}^{2p}}
\end{align}
With $\brhowithin$ fully specified, we generate the synthetic datasets $\hat \brho_{i,k} = \rho(\phi_{\rm LS},\T) + \boldsymbol {\omega}_{i,k}$ using a pseudorandom number generator.  Evaluation of $\H[\hat \brho_{i,k},\T]$ results in an empirical distribution for $\brhowithin$, from which we compute $\TgBari$ and its variance.

\begin{figure*}
\begin{center}
\hspace*{-5mm}\includegraphics[width=16cm]{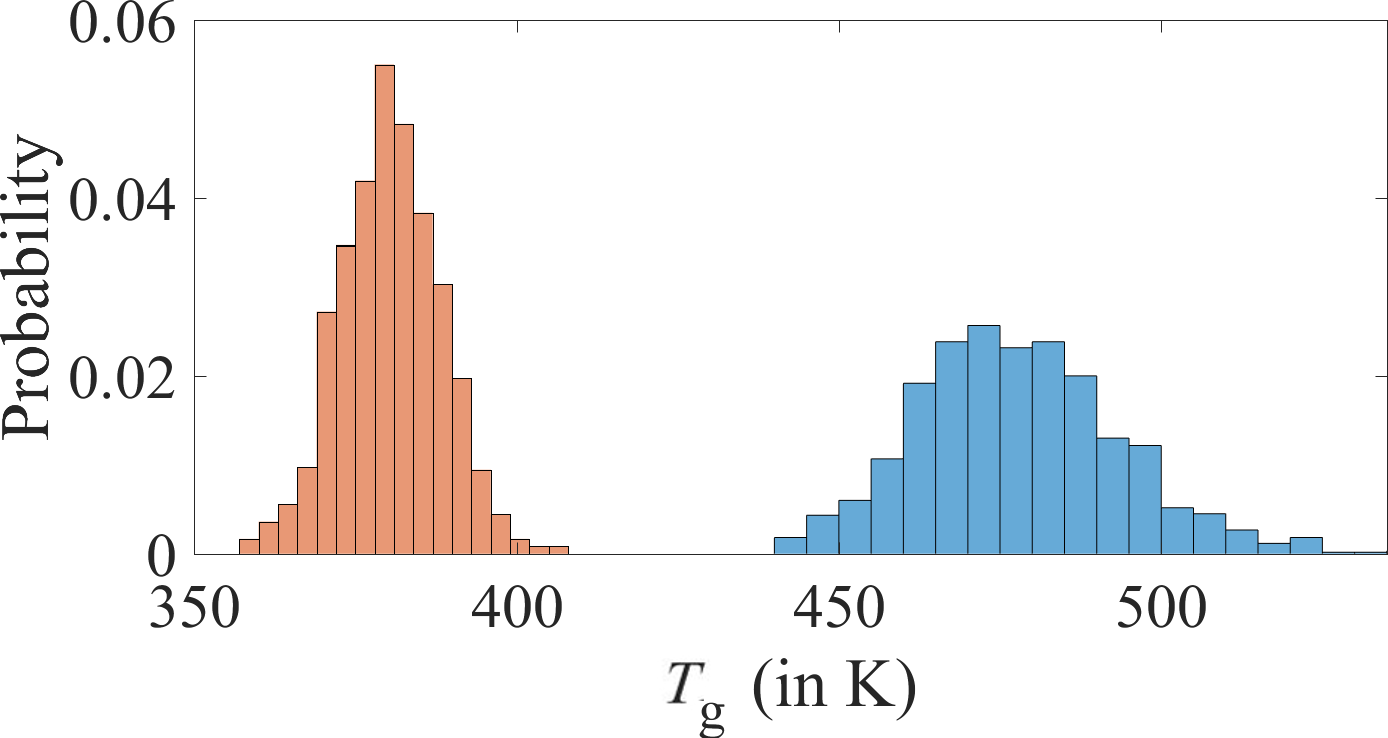}
\caption{Histograms of \Tg values from the hyperbola method applied to synthetic data generated from two independent 3800-atom 33BF datasets. \changeAD{These are different simulations of the same chemistry.}  The lack of overlap between the distributions suggests that \changeAD{$\tgwithin$, i.e., the uncertainty resulting from propagation noise in $\rho(T)$ data, does not capture all of the uncertainty in \tg.}}\label{fig:noverlap} 
\end{center}
\end{figure*}

 Figure~\ref{fig:noverlap} shows two histograms computed from two different 33BF datasets using 1200 synthetic datasets each.  The lack of overlap in the distributions is yet another indicator that the between-simulation noise discussed in Eq.~\eqref{eq:mixedeffects} has a significant effect on the uncertainty in simulated \Tg estimates.

\subsubsection*{Between uncertainty and weighted-mean averages}
\label{subsec:WeightedMean}
In the previous section, we presented a noise-propagation method for estimating the within-simulation uncertainty $\tgwithin$.  It remains to determine the between uncertainty $\tgbetween$, and to compile all simulated results for \Tg into a single estimate with its associated (combined) uncertainty.

Denote by $\TTg=(\hat T_{{\rm g},1},\ldots,\hat T_{{\rm g},M})$ the collection of $M$ simulation-based estimates of \tg.  Given our statistical model for \tg, we postulate that
\begin{subequations} 
\begin{align}
\tgwithin &=\mathcal N(0,\varsigma_i^2)\\
\tgbetween&=\mathcal N(0,y^2)
\end{align}
\end{subequations}
where $\varsigma_i$ depends on the $i$th dataset and $y$ is a constant variance that accounts for affine transformations and rotations apparent in Fig.~\ref{fig:noverlap1}.  Given this, the joint probability density function for $\TTg$ is given by
\begin{align}
f(\TTg|\bar\tg,y) 
= \frac{\exp\left(-\frac12\sum_i\frac{(\TgBari-\bar\tg )^2}{\varsigma_i^2+y^2}\right)}
       {(2\pi)^{M/2}\sqrt{\prod_i(\varsigma_i^2+y^2)}}   \label{eq:likelihood}
\end{align}
Considering the above as a function of $(\bar\tg,y)$ and maximizing the log-likelihood {[i.e.\ the logarithm of Eq.~\eqref{eq:likelihood}]} results in the equations
\begin{subequations}
\begin{align}
\displaystyle \sum_{i=1}^M \frac{\TgBari-\bar\tg}{\varsigma_i^2+y^2} &= 0\\
\label{eq:mleBetweenVariance}
\displaystyle \sum_{i=1}^M \frac{1}{\varsigma_i^2+y^2} - \sum_{i=1}^M \frac{(\TgBari-\bar\tg)^2}{(\varsigma_i^2+y^2)^2} &= 0
\end{align}
\end{subequations}
 for a critical point.  Solution to the first \changeAD{results in estimating \Tg by the} weighted average
\begin{align}
\mathcal T = \left[\sum_{i=1}^M \frac{1}{y^2 + \varsigma_i^2} \right]^{-1}\sum_{i=1}^M \frac{\TgBari}{y^2 + \varsigma^2_i} \label{eq:fullweightedmean}
\end{align}
Setting $y=0$ results in the usual minimum variance estimator of \Tg as the mean of independent, \changeAD{random variables, each with different variance}.  However, as is well-known, this estimation suffers from over-emphasizing experiments (or computations in our case), whose uncertainty $\varsigma_i^2$ is underestimated.  The central benefits of Eq.~\eqref{eq:fullweightedmean} are that it accommodates additional, unaccounted for uncertainties, referred to colorfully as dark-uncertainties.\cite{Rukhin}  In doing so, Eq.~\eqref{eq:fullweightedmean} also avoids up-weighting overconfident results.  Note that when $\varsigma_i^2 \ll y^2$ for all $i$, Eq.~\eqref{eq:fullweightedmean} returns the sample average; i.e., the estimator views each realization $T_i$ as an equally meaningful draw from a statistical ensemble.    For more details on Eq.~\eqref{eq:fullweightedmean} we refer to Ref.\ \ocite{Rukhin} and its many references.

It remains to determine the quantity $y$.  Several approaches have been recommended; see Ref.~\ocite{Rukhin} for a review.  In our case, we complete the maximum likelihood analysis by substituting the weighted mean Eq.~\eqref{eq:fullweightedmean} into Eq.~\eqref{eq:mleBetweenVariance}, and solving the non-linear system for $y$. Having determined the model parameters $\mathcal T$, $y$, and $\varsigma_i$, Ref.~\ocite{Rukhin} provides a consistent estimator for the variance of $\mathcal T$ having the form
\begin{align}
\delta^2 = \left[ \sum_{i=1}^M \frac{1}{y^2+\varsigma_i^2} \right]^{-2} \sum_{i=1}^M \frac{(\TgBari- \mathcal T)^2}{(y^2+\varsigma_i^2)^2}
\end{align}
Notably, $\delta$ has the property that $\delta \to 0$ as the number of simulations $M\to \infty$, provided the $\varsigma_i$ are well behaved.  We use $\delta$ to indicate the confidence with which we have determined the true \Tg value predicted by arbitrarily large MD simulations of a given chemistry.  That is, we approximate the  99\% confidence interval for \Tg to be $\mathcal T \pm 3{\delta}$. 

\begin{figure*}\begin{center}
\includegraphics[width=16cm]{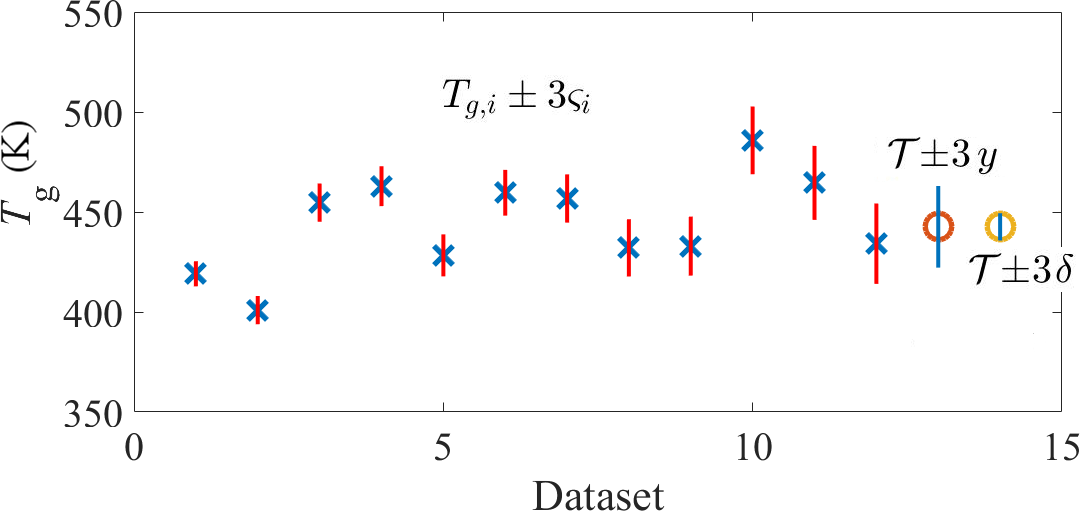}\caption{33BF \Tg estimates from individual datasets and according to Eq.~\eqref{eq:fullweightedmean}.  Blue $\times$ indicate hyperbola analyses applied to individual datasets; red error bars correspond to $3\varsigma_i$ values of the within-simulation uncertainty.   The orange o corresponds to \Tg estimates $\mathcal T$ via Eq.~\eqref{eq:fullweightedmean}.  The corresponding blue error bar (second from right) corresponds to $\pm 3 {y}$, illustrating that the between-simulation uncertainty $y$ accounts for finite-size and finite-time effects between individual realizations.  The short error bar on the far right corresponds to $\pm 3 \delta$.  It is small by virtue of the fact that we reduce uncertainty by combining results from many simulations. }\label{fig:errorbars}\end{center}
\end{figure*}

In Fig.~\ref{fig:errorbars}, we show $T_g$ estimates for individual 33BF datasets according to Eq.~\eqref{eq:fullweightedmean}.   Note that many datasets yield predictions whose error bars fail to overlap.  The between-simulation uncertainty is shown on the right side of the plot and illustrates that $y$ accounts for this lack of consistency between individual simulations.  We also plot the 99\% confidence interval $\mathcal T \pm 3\delta$, which is small by virtue of the fact that $M$ is relatively large.

\section*{Concluding Thoughts}
\label{sec:conclusion}

Having gone through the analysis and tutorials presented in this chapter, the reader has hopefully come to appreciate that uncertainty quantification of molecular dynamics is a surprisingly rich and often complicated endeavor.  In looking back, we revisit two themes of the discussion.

First, uncertainty quantification ultimately aims to facilitate decision making.  When possible, \changeAD{one} should therefore balance the thoroughness of analyses against the weight of the decision to be made.  We saw this, for example, in the tutorial on ensemble verification when modeling the PDFs required in Eq.~\eqref{eq:shirtslinear}.  In principle, we could have invoked a variety of UQ tools to better quantify the agreement with theory, given that uncertainty in the mode weights $a_m$ can be estimated via Eq.~\eqref{eq:standerr}.  However, as we posed it, this verification test only provides qualitative information about the simulation, and therefore does not require the full arsenal of available tools.  In contrast to this, the \tg ~analysis was more thorough in its modeling of the within and between uncertainties.  Motivating this, however, is the \changeAD{fact} that such simulations are routinely used by industrial scientists to guide \changeAD{new materials} development.  Ultimately such development requires a significant investment in experimental resources to validate the simulated findings and satisfy regulatory requirements.  As such, unverified and unreliable simulations can lead to significant monetary loss if used as the basis for such decisions. \changeAD{The consequence of under-informed decisions in this case justifies the thoroughness of the analysis.}

A second theme that has hopefully become apparent is that UQ of MD cannot be treated independently of the specifics of molecular dynamics.  The trajectory analysis tutorial  had at its heart a fundamental question: does MD represent the trajectory of a Hamiltonian system?  Moreover, in order to address this question, we had to dive headlong into the symplectic structure of Hamilton's equations, an area that constitutes ongoing and active research in the field of dynamical systems.  Furthermore, the other tutorials required input from physics, such as the ensemble verification work by Shirts \cite{Shirts13} or the observation that bilinear behavior is a characteristic feature of acceptable \tg ~data.  

In light of these latter examples, we therefore wish to emphasize that UQ of MD can benefit from scientists \changeAD{at all stages of the data generation and analysis pipeline.}
This is illustrated perhaps most clearly by the hyperbola fit of density-temperature data, which is neither a complicated idea nor a widely used tool\blfootnote{Incidentally, we independently conceived of the idea of using hyperbola fits to model bilinear data, only to discover that this had been proposed several decades earlier in Ref.~\ocite{Cologne}.}.  Moreover, one need not know anything about MLE or Bayesian statistics to recognize that \changepp{hyperbolas} can be useful for identifying asymptotic regimes, as required by the physics in question.  In fact, we take a somewhat opposing perspective, arguing that hardened statistical tools such as MLE do not reach their full potential without the insight and creativity afforded only to modelers, who know their own data best.  With that in mind, we emphasize that UQ of MD is not a collection of tools to be used only by experts, but rather a continually growing field that requires the input and experience of end-users.  

In this light, we wish to remind the reader that many of the most important and open problems are ones that we have only lightly touched upon.  Calibration and uncertainty quantification of force fields remains one of the most challenging, owing the range of assumptions and approximations that underpin their construction.  A fuller understanding of the symplectic structure of discrete integrators and implications for sampling remains another area in which progress is needed. Perhaps more practically, however, the community also requires a wider adoption of those techniques that are well known. \changeAD{In summary, ongoing research efforts in uncertainty quantification combined with increased diligence in their application, are essential to capture the full benefits of molecular dynamics simulation for materials development.}

\bibliography{chapter_references_f}
\bibliographystyle{revcompchemj_3}

\end{document}